\documentclass[12pt,reqno]{article}
\usepackage[english]{babel}
\usepackage{amsmath,amsthm,commath,mathrsfs,amssymb,extarrows}
\usepackage{dsfont}

\usepackage{subfigure}  %%% Add command

\usepackage{tikz}
\usetikzlibrary{arrows,intersections}

\usepackage[colorlinks = true,urlcolor = blue,citecolor = blue,breaklinks]{hyperref}
\usepackage{amsfonts}
\usepackage{graphicx}
\usepackage{enumerate}

\usepackage[right,pagewise,displaymath, mathlines]{lineno}
\usepackage{epstopdf}
\usepackage{color}
\usepackage{multirow}
\usepackage{authblk}
\usepackage[round]{natbib}
\usepackage{float}
\restylefloat{figure,table}

\usepackage{colortbl}

\usepackage{geometry}
\numberwithin{equation}{section}
\numberwithin{figure}{section}
\numberwithin{table}{section}

\theoremstyle{definition}

\newtheorem{note}{Note}[section]

\numberwithin{equation}{section}

\usepackage{caption}

\definecolor{darkred}{rgb}{0.7, 0, 0}
\definecolor{darkbrown}{rgb}{0.55, 0.2, 0.15}
\definecolor{darkblue}{rgb}{0.1,0.1,0.6}
\definecolor{darkgreen}{rgb}{0.1,0.5,0.2}

\usepackage{xr}
\makeatletter

\newcommand*{\addFileDependency}[1]{\typeout{(#1)}
\@addtofilelist{#1}
\IfFileExists{#1}{}{\typeout{No file #1.}}
}\makeatother

\newcommand*{\myexternaldocument}[1]{%
\externaldocument{#1}%
\addFileDependency{#1.tex}%
\addFileDependency{#1.aux}%
}

\myexternaldocument{NLP-revision-2-supplement} %%% FILE NAME

\title{\Large\bf NLP-based detection of systematic anomalies among the narratives of consumer complaints}

\author[1]{Peiheng Gao}

\author[2]{Ning Sun}

\author[3]{Xuefeng Wang}

\author[,4,5]{\break Chen Yang\thanks{Corresponding author Chen Yang, PhD, ASA; e-mail \href{mailto:chen.yang@mountsinai.org}{chen.yang@mountsinai.org}}}

\author[1]{Ri\v cardas Zitikis}

\affil[1]{\normalsize School of Mathematical and Statistical Sciences, Western University, London, Ontario~N6A~3K7, Canada}

\affil[2]{\normalsize{Agri-Food Analytics Lab, Dalhousie University, Halifax, Nova Scotia~B3H~4R2, Canada}}

\affil[3]{\normalsize Department of Financial Risk Analytics and Reporting, Canada Life, Toronto, Ontario~M5G~1R8, Canada}

\affil[4]{\normalsize Department of Population Health Science and Policy, Icahn School of Medicine at Mount Sinai, New York, New York~10029, USA}

\affil[5]{\normalsize Institute for Health Care Delivery Science, Icahn School of Medicine at Mount Sinai, New York, New York~10029, USA}

\date{}

\begin{document}
%\linenumbers

\maketitle

\begin{center}\large 
  To appear in \href{https://www.risk.net/journal-of-operational-risk}{\it Journal of Operational Risk}
\end{center}

\newpage 

\noindent 
\textbf{Abstract.}
A Natural Language Processing (NLP) based procedure  is developed for detecting systematic non-meritorious consumer complaints, simply called systematic anomalies, among complaint narratives. While classification algorithms are used to   identify meritorious complaints, in the case of smaller and frequent systematic   patterns of non-meritorious complaints, those algorithms may falter due to a variety of reasons, including technical ones as well as natural limitations of human analysts. Therefore, at the next stage after classification, the complaint narratives   are converted into quantitative data,  which are then analyzed using   indices for detecting systematic anomalies.  An illustration of the entire procedure   is provided using complaint narratives from the Consumer Complaint Database of the Consumer Financial Protection Bureau.  The results suggest that the Support Vector Machine (SVM) outperforms other selected classifiers. Although classification results with the VADER intensity pertinent to the featurization step have lower accuracy, they contain fewer non-meritorious complaints than those without the VADER intensity.

\bigskip


\newpage 

\section{Introduction}
\label{intro}

The narratives of consumer complaints contain valuable information about the sales and underwriting activities in the insurance industry. Responding to consumer complaints is one of the primary ways for insurance departments \citep{naic2023} to regulate market practices \citep{k1995}. In 2021, state insurance departments received 259\,345 official complaints, among which 1\,474 market conduct examinations were triggered and completed \citep{naic2022}. By analyzing consumer complaint data, regulators can identify market mis-behaviour, such as false sales illustrations or failure to pay legitimate claims on a timely basis, and initiate market conduct examinations to determine if disciplinary procedures are needed.  

In addition to regulatory purposes, consumer complaints also indicate service quality. For example, a survey about customer service of 140 companies from 14 industries has shown that the three dominating worst-rated industries are communications, banking/financial services, and insurance \citep{wm2010}. As a result, bulk complaint datasets, such as the one by the Consumer Financial Protection Bureau \citep{CFPB}, have been created.

Certainly, datasets such as the CFPB consumer complaints also serve the purpose of regulation, as using complaints for regulatory purposes is significantly more cost-effective than resolving them individually  \citep{l2014}. Analysts have attempted to determine whether complaints have merits and possibly constitute violations of state laws or regulations \citep{k1995}. In such cases, companies may voluntarily provide reliefs along with responses to the complaints. These reliefs can be seen as risk premiums to mitigate regulatory or litigation risks in the cases described by complaint  narratives. Hence, reliefs, particularly the tangible ones, are critical indicators of potential violations of state laws or regulations. Among the complaints in the CFPB database, companies have granted consumers tangible reliefs 21.9\% of the time.  In the remaining cases, the focus has been on providing consumers with explanations \citep{l2014}. Nevertheless, whether intentionally or not, companies are reluctant to grant reliefs in some cases, which  are viewed in the present paper as anomalies  of the dispute-resolving system.  

\subsection{Motivation}

Reliefs to consumer complaints should be granted with prudence. From the economic perspective, reliefs to consumer complaints are costly from the time, effort, and monetary perspectives, and hence impact the profitability of businesses. Moreover, high percentages of reliefs granted to complaints suggest poor levels of compliance with regulations or operational risk management, and hence influence the reputation of financial institutions by attracting fewer customers although more media attention and regulation. For this reason, financial institutions need to
implement operational risk management measures to control the percentage of
complaints that deserve reliefs. Indeed, taking risks brings opportunities, and so completely avoiding risks to minimize the aforementioned percentage may not be attractive from the business perspective. Yet, the percentage cannot be too high due to reasons such as arising reputational and regulatory issues. 

With the help of records of monetary and non-monetary reliefs granted to complaints provided by the publicly available CFPB dataset, machine learning based classifiers can be trained to assist in determining whether reliefs should be granted. Among the complaints submitted to the CFPB, those that are closed with monetary or non-monetary reliefs are defined as ``meritorious'' \citep{HJP2021} throughout the present paper . 

There are several reasons to suggest that more effort is needed for this classification task. First, unlike typical classification tasks of text inputs, classifiers cannot be trained by using comments along with polarities (e.g., tweets) or the number of stars (e.g., online shopping experiences) as responses. Second, complaints are submitted by consumers whereas decisions to grant, or not to grant, reliefs are made by  specialists of financial institutions. As a result, the potential features extracted from the narratives are often less correlated with the responses, thus leading to relatively low accuracy if compared with the polarity classification of tweets or online shopping experiences. Third, extra steps are needed to check if more complaints can actually be filtered out from the pool of meritorious complaints.

\subsection{Related work}

For those datasets that include the narratives of complaints, the Natural Language Processing (NLP) based analyses have become popular. For example, \citet{kp2018} propose a computational approach for analyzing  complaints from   GEICO, an insurance company with more than 16 million policies and 24 million insured vehicles. However, only the Latent Dirichlet Allocation (LDA) based topic analysis is performed such that 30 categories of topics are identified without any direct relationship to decisions of pay-offs or reliefs. Using a similar method, \citet{LCKND2020}   consider the LDA-based topic analysis for an insurance company of personal lines to improve its customer services by analyzing customer calls.  As the LDA-based topic analysis is an unsupervised learning technique, this study does not use decisions related to payoffs or reliefs as classification labels, although it involves other NLP techniques such as text featurization and sentiment analysis. Furthermore, based on the CFPB dataset, \citet{os2022} investigate the impact of a major banking scandal using the Valence Aware Dictionary for sEntiment Reasoning (VADER) \citep{HG2014}.  However, this study directly quantifies text inputs using VADER without a further exploration of text mining or classification problems. \citet{PFMK2022} provide an overview of how machine learning helps to categorize textual descriptions of operational loss events into  the seven Basel II event types  of operational risk by applying PYTHON implementations of the SVM and multinomial naive Bayes algorithms to pre-categorized \"{O}ffentliche Schadenf\"{a}lle OpRisk (\"{O}ffSchOR) data.

In the present study, the inputs are the textual descriptions of operational loss events \citep[e.g.,][]{vgpz2023}, which have higher quality due to  professional narratives compared with consumer complaints that contain more typos but less informative details. \citet{WCL2022} introduce a text mining method to analyze changes in operational risk in an innovative way based on a massive data set of textual risk disclosures of financial institutions, which aggregate risk perceptions of senior managers of the financial industry.  However, this study employs textual data analytic techniques to identify significant operational risk drivers instead of performing machine learning tasks such as classification. \citet{JDPCP2023} develop a risk-based knowledge graph to integrate text mining and analytic hierarchy process risk assessment with knowledge graphs for operational risk analysis.  While text mining techniques are integrated in this study, quantifying operational risk rather than assisting decision-making is still the ultimate goal  \citep[e.g.,][]{WCL2022,os2022}. 
\citet{vgpz2024} provide a tweet data analysis for detecting emerging operational risks.

\subsection{Research gaps}

Due to poor quality of narratives as textual data inputs as well as weak correlation between the inputs (from customers) and outputs (from financial institutions), the existing text classification methods have shortcomings, particularly from the perspective of limiting the percentage of meritorious complaints. These considerations have motivated us to work on improving classification procedures from several perspectives, including the featurization of textual data inputs, and to also propose additional analyses using an anomaly detection method to assess if the resulting meritorious classes actually contain systematic non-meritorious complaints.

\subsection{The scope}

In the present paper, a systematic-anomaly detection procedure  is proposed to assess complaints for which companies are reluctant to provide reliefs. In the current study, the VADER sentiment intensities play an important role in the text classifier training stage as well as in the performance evaluation stage. In the text classifier stage, the VADER sentiment intensities are combined with the Term Frequency-Inverse Document Frequency (TF-IDF) for the purpose of feature extraction. In the performance evaluation stage, the VADER sentiment intensities are combined with the discounted dollar amounts mentioned in the complaint narratives as well as with the word counts of the cleaned narratives. This, in turn, facilitates the use of anomaly detection indices, which test if there are persistent anomalies in the dataset of complaints. To illustrate the proposed methodology, we use the CFPB dataset, due to its transparency and accessibility.

The present study is comprised of several stages. At the classification stage, the labels are taken directly from the CFPB complaint dataset, with the complaints that were resolved by either monetary or non-monetary reliefs labeled as ``meritorious,'' while complaints with only responses labeled as ``non-meritorious'' \citep{HJP2021}. The selected complaints are then restricted to those consumer complaints that involve specific dollar amounts related to personal credit or debit card accounts against a selected major financial institution, specifically, the Bank of America (BOA). The features of the classification task are extracted from the complaint narratives pre-processed by NLP and transformed to numerical values using the widely utilized TF-IDF method and our modified version of the TD-IDF using the VADER sentiment density. The classifiers considered in this study include the Logistic Regression (LR), Support Vector Machine (SVM), Gradient Boosting (GB), Multilayer Perceptron (MLP), and Random Forest (RF), which are widely used in text classification studies \citep[e.g.,][]{H2020}. At the performance evaluation stage, the resulting meritorious class is further assessed by the index for systematic-anomaly detection \citep{GZ2018,GZ2019a,GZ2019b,GZ2020}.

\subsection{Contributions}

The present study has several important implications on how financial institutions can better utilize massive volumes of customer complaints in order to train classifiers that support decision making when resolving disputes. Among the achievements are: 
\begin{enumerate}[(1)]
\item 
Exploring the possibility of improving textual data classification by incorporating the VADER sentiment density. 
\item 
Introducing an anomaly-detection method that controls the percentage of reliefs granted to complaints. 
\item 
Evaluating classifiers by examining the extent at which the resulting meritorious class includes actually non-meritorious complaints in a systematic pattern. 
\item 
Demonstrating the usefulness of the proposed procedure in identification of meritorious customer complaints in the CFPB dataset. 
\end{enumerate}

The rest of the paper is organized as follows. 
In Section~\ref{DD}, a subset of the CFPB consumer complaint data   is chosen for illustration of the proposed procedure. In Section~\ref{sec: classification},  selected classification algorithms are employed for anomaly detection among complaint narratives. 
In Section~\ref{sec: sentiment score}, the narratives  are converted into quantitative data,  and a basis for an input-output system  is established. 
In Section~\ref{sec: input-output}, two input-output systems arising from two featurization methods are introduced.  
In Section~\ref{sec: performance}, anomaly detection indices based on the two input-output systems  are introduced, and their ability to detect anomalies is evaluated.  
Section~\ref{sec: conclusion} concludes the paper.
Two appendices  illustrating the steps of corpus cleaning \citep[e.g.,][]{vgpz2023} as well as the behaviour of the $B$-index with respect to increasing sample sizes are in the online supplement to this paper.

\section{Data}
\label{DD}

A subset of complaint narratives from the Consumer Complaint Database of the Consumer Financial Protection Bureau \citep{CFPB} is selected for illustrative purposes. Extracted are the records of those complaints whose narratives      
\begin{enumerate}[(1)] 
\item 
are related to credit cards and prepaid cards of a major bank, 
\item 
were submitted between December 1, 2011, and June 29, 2023.  
\end{enumerate}
Of all the selected complaint records, only those referring to unique dollar amounts are considered. Furthermore, the complaints referring to dollar amounts exceeding \$10\,000   are excluded because those complaints are more likely to be associated with business accounts rather than individual accounts. Hence, the resulting data consist of 2\,905 records of complaints related to credit card and prepaid card services of the bank. For simplicity, complaints with only single positive dollar amounts appearing in the narratives   are selected. As a consequence, 2\,849 complaints  are ultimately selected for the current study.

The complaints closed without any monetary or non-monetary reliefs are viewed as non-meritorious \citep{HJP2021} and called anomalies throughout the present paper.  Among the selected 2\,849 complaints, 1\,051 are meritorious, that is, they were resolved with some reliefs. As there are dollar amounts submitted at different times, we discount the amounts using the annual  Consumer Price Index (CPI), which is defined \citep{CPI}0
\[
\text{Year 2 Price} = \text{Year 1 Price}\times(\text{Year 2 CPI} / \text{Year 1 CPI})
\]
to the base date of January 1, 2015. Table~\ref{Tab: summary} 
\begin{table}[h!]
\centering
\begin{tabular}{llrl}
\hline\hline 
Date &  Raw complaint narratives  & Adjusted & Merit  \\
\hline
11/10/2016 & Macys did not reverse out my \$230.00 \dots & \$227.53 & Yes 
\\
11/28/2017 & Inquired about \$300.00 increase \dots & \$287.36 & No 
\\
04/14/2022 & A charge of \$170.00 was made \dots & \$107.15 & No 
\\
06/12/2023 & Someone fraudulently charged \$750.00 \dots & \$360.83 & Yes \\
\hline 
\end{tabular}
\caption{Narrative excerpts, adjusted dollar amounts, and merit indicators.}
\label{Tab: summary}
\end{table}
illustrates.

\section{Anomaly detection via classification}
\label{sec: classification}

As all non-meritorious complaints are seen as anomalies throughout the present paper, classifiers are trained to identify the complaints as either meritorious or anomalies. To this end, appropriate feature extraction is needed to translate the cleaned texts into a numerical matrix.

\subsection{Featurizations}

Anomalies, which are non-meritorious complaints in the context of the present paper, are identified through classification.  To train the classifiers, two featurization  methods are considered  for text inputs. The first one is the Term Frequency--Inverse Document Frequency (TF-IDF),  which we recall with adaptations to our current research while at the same time introducing the necessary notation. 

Namely, let $\mathcal{D}=\{d_1,\ldots,d_n\}$ denote the set of all cleaned complaint narratives and $\mathcal{W}$ the set of all words appearing in $\mathcal{D}$. Then the Term Frequency $\text{TF}:\mathcal{D}\times\mathcal{W}\mapsto\mathbb{N}$ is defined  \citep{s1972} as the frequency of the $j^{\text{th}}$ word $w_j$ ($j=1,\ldots,\lvert\mathcal{W}\rvert$) in the $i^{\text{th}}$ cleaned complaint narrative $d_i$ ($i=1,\ldots,n$), where $\lvert\mathcal{W}\rvert$ denotes the number of elements in the set $\mathcal{W}$. The Inverse Document Frequency   $\text{IDF}: \mathcal{W}\mapsto[0,\infty)$ is defined as
$$\text{IDF}(w_j) = \log\bigg(\frac{n}{n_j}\bigg)$$
for $w_j\in\mathcal{W}$, where 
$$n_j = \sum^n_{i=1}\mathds{1}\{\text{TF}(d_i, w_j) > 0\}, $$ 
with $\mathds{1}$ denoting the indicator, that is, $n_j$ is the number of cleaned narratives that include $w_j$. Then $\text{TF-IDF}: \mathcal{D}\times\mathcal{W}\mapsto[0,\infty)$ is defined by the equation  \citep{blg1998} 
$$\text{TF-IDF}(d_i, w_j) = \text{TF}(d_i, w_j)\times \text{IDF}(w_j).$$
Using the TF-IDF featurization, the cleaned complaint narratives are translated into a matrix with $n$ rows and $\lvert \mathcal{W}\rvert$ columns.

The second featurization method is the TF-IDF-VADER, which is based on the function  $\text{TF-IDF-VADER}: \mathcal{D}\times\mathcal{W}\mapsto[0,\infty)$ defined by
$$\text{TF-IDF-VADER}(d_i, w_j) = \text{TF-IDF}(d_i, w_j)\times \lvert \text{VADER}(w_j)\rvert ,$$
where $\text{VADER}(w_j)$ denotes the VADER sentiment intensity of the word $w_j$ \citep{HG2014}.  
Since   the corpus consists of complaints    which have negative polarities, all the words with non-negative VADER sentiment intensities   are discarded from the Bag of Words. Hence the TF-IDF-VADER approach is applied on only those words that belong to 
\[
\mathcal{W}_-:=\left\{w\in\mathcal{W}: \text{VADER}(w) < 0\right\}, 
\]
thus giving rise to a featurized matrix with $n$ rows and $\lvert \mathcal{W}_-\rvert$ columns.

\subsection{Classifiers}

When a new segmentation of classes is defined, or a new method is proposed, the selected classifiers are usually tested in the exploratory stage in the hope of maximizing the performance of the proposed method. For example, to classify ``facts'' and ``opinions'' from tweets, \citet{cdlsj2018} have selected the LR, SVM, RF, extreme GB, and Deep Neural Network for exploratory purposes. Another example is \citet{kv2022} who, at the exploratory stage, tested oversampling methods for the issue of imbalanced classes using the LR, SVM, RF, and Na\"{i}ve Bayes.

To the best of our knowledge, this present paper is the first one to attempt to classify ``meritorious'' and ``non-meritorious'' complaints from the CFPB database. Five classification methods  are selected for the exploratory stage: the LR, SVM, GB, MLP and RF, which are a subcollection of those used by \citet{H2020}. For each of these classification methods, the selected 2\,849 complaint narratives  are split randomly so that roughly 60\% make up the training set and the remaining ones make up the validation set. By repeating this procedure 500 times, 500 confusion matrices for each of the five classifiers   are obtained using the TF-IDF and TF-IDF-VADER methods. The average classification accuracy (i.e., the percentage of correct predictions regardless of the true label), the percentage of correctly classified meritorious complaints, and the F1 score based on the 500 confusion matrices for each of the five classifiers are reported in Table~\ref{accuracy-result-vader}.
\begin{table}[h!]
\centering
\begin{tabular}{lccccccc}
\hline\hline 
\multicolumn{1}{c}{Model} & 
\multicolumn{3}{c}{TF-IDF} & 
\multicolumn{1}{c}{} & 
\multicolumn{3}{c}{TF-IDF-VADER} 
\\ 
\cline{2-4}  \cline{6-8}  
    & Accuracy & Merit   &    F1     & & Accuracy & Merit   &    F1 \\ \hline 
LR  & 67.37\%  & 15.41\% &   78.00\% & & 63.96\%  & ~~8.52\% &    76.80\% \\ 
SVM & 67.30\%  & 13.24\% &   78.24\% & & 63.72\%  & ~~7.29\% &    76.65\% \\ 
GB  & 64.61\%  & 16.43\% &  76.03\%  & & 62.75\%  & 11.31\% &    75.44\% \\
MLP & 58.08\%  & 36.04\% &  71.22\%  & & 56.50\%  & 34.96\% &    66.38\% \\ 
RF  & 65.56\%  & ~~9.80\% &  77.64\%  & & 60.67\%  & 18.71\% &   73.28\% \\ 
\hline
\end{tabular}
\caption{Average classification accuracy, merit, and F1 scores for 500 random splits of training and testing sets under the TF-IDF and  TF-IDF-VADER featurizations.}
\label{accuracy-result-vader}
\end{table}

Based on Table~\ref{accuracy-result-vader}, it can be concluded that the TF-IDF featurization consistently generates higher accuracy than the TF-IDF-VADER featurization.  Moreover, under the TF-IDF, the SVM slightly outperforms the other four classifiers, as seen from the accuracies and F1 scores, while the LR SVM slightly outperforms the other four classifiers under the TF-IDF-VADER featurization. In addition, fewer complaints are classified as meritorious under the TF-IDF-VADER than under TF-IDF for all classifiers except for the RF. In summary, the VADER sentiment density does not seem to help to improve the classification accuracy or the F1 score. However, it helps to control the percentage of complaints that are resolved with reliefs.

The average accuracy of the classification of meritorious versus non-meritorious complaints is between 60\%-70\%, which is fairly low if compared with some other studies. However, such classification performance is not rare when sentiment analysis is involved, as is the case with, for example, the sentiment classification of the IMDB data \citep[Table 1]{O2023}. We see from the latter study that the MLP is outperformed by the LR, SVM, and RF in both the sentiment classification of the IMDB data and the PadChest data \citep[Table 4]{O2023}.

\section{Converting narratives into quantitative data}
\label{sec: sentiment score}

As our input-output system requires numerical inputs and outputs, quantities related to the cleaned meritorious narratives need to be identified for the purpose of anomaly detection. In this section, the word counts of the cleaned narratives, discounted dollar amounts (obtained via the CPI as described earlier), and VADER sentiment intensities \citep{HG2014} are chosen for constructing input-output systems based on several economic considerations.

\subsection{TF-IDF}

To measure the sentiment level of the CFPB consumer complaints,   the function $\textrm{VADER}:\Omega \to [-4,4] $   is employed with $\Omega $ denoting the Bag of Words of the VADER sentiment lexicon that has been widely used in finance \citep[e.g.,][]{os2022, llxy2023}, public health \citep[e.g.,][]{s2021}, and other areas. Namely, for each consumer $i$, the sentiment score  is defined by
\begin{equation}\label{def-si}
s_i=\sum_{w_j \textrm{ in } d_j \textrm{ and } W_{-}} \big| \textrm{VADER}(w_j) \big|, 
\end{equation}
where the sum is taken over all words $w_j $ that are both in the cleaned narrative $d_i$ and in the set $ W_{-}$ of those words that have negative VADER sentiment intensities. Hence, the sentiment score $s_i$ can be rewritten as 
\[
s_i= \sum_{w_j \textrm{ in } d_j}   \mathds{1}\{ \textrm{VADER}(w_j)<0\} \big(-\textrm{VADER}(w_j) \big),
\]
where $\mathds{1}$ denotes the indicator function. It is instructive to compare this sentiment score with the one used by \citet{os2022}.

The time and effort invested by the consumer $i$ to generate a complaint depend on 1) how much money the consumer was charged or how much return the consumer has not received, and 2) the consumer's sentiment level due to  monetary or non-monetary losses. With the number $m_i^{\textrm{TI}}$ of words in the cleaned narrative $d_i$, the adjusted  dollar amount $l_i$, and the sentiment score $s_i$, we assume 
the Cobb-Douglas relationship 
\begin{equation}\label{CD-0}
m_i^{\textrm{TI}} = \beta s_i^{\alpha}l_i^{1-\alpha} , 
\end{equation}
where $m_i^{\textrm{TI}}$ is viewed as the yield, $l_i$ and $s_i$ are viewed as factors, 
and $\alpha\in(0,1)$ and $\beta>0$ are the geometric weight and the scale, respectively. Note that by the Cobb-Douglas relationship, the law of diminishing marginal utility is satisfied for the sentiment score $s_i$ and the adjusted dollar amount $l_i$. 

\begin{note}\label{vader-4-0}
Since $\textrm{VADER}(w)\in [-4,4]$ for every word $w$ \citep{HG2014}, the sentiment score $s_i$ is non-negative and does not exceed $4m_i^{\textrm{TI}}$. Hence,  $s_i/(4m_i^{\textrm{TI}}) \in [0,1]$. 
\end{note}

\begin{figure}[h!]
\centering
\includegraphics[width=0.5\textwidth]{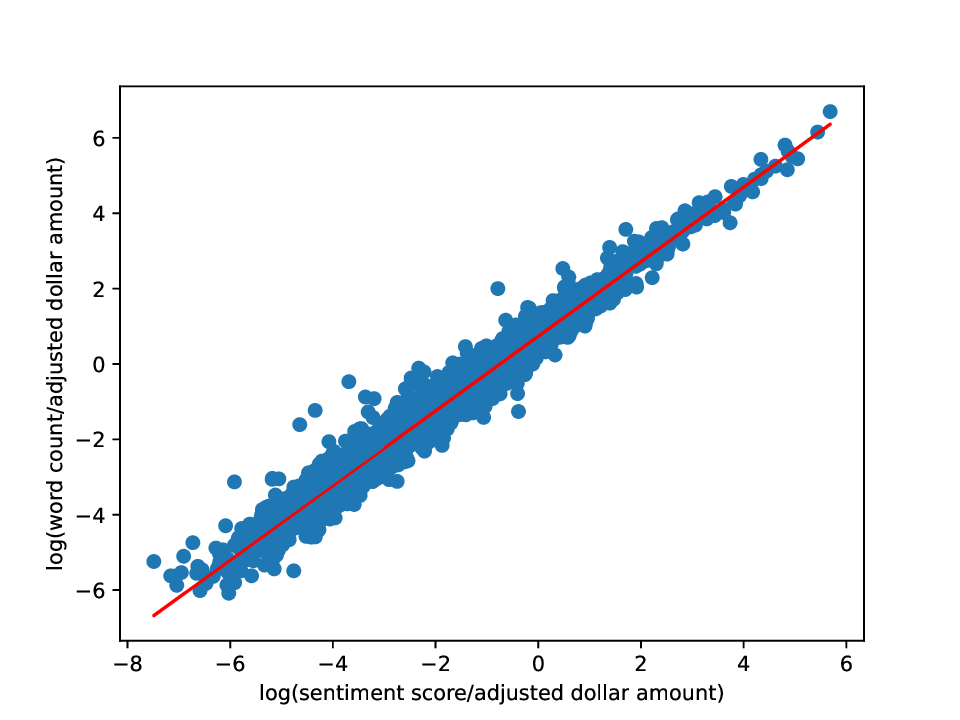}
\caption{Scatter-plot~\eqref{log CD-0sp} and  the fitted least squares regression line.}
\label{CM2-0}
\end{figure}

\begin{figure}[h!]
\centering
\subfigure[Residuals vs Fitted]   {\includegraphics[width=0.49\textwidth]{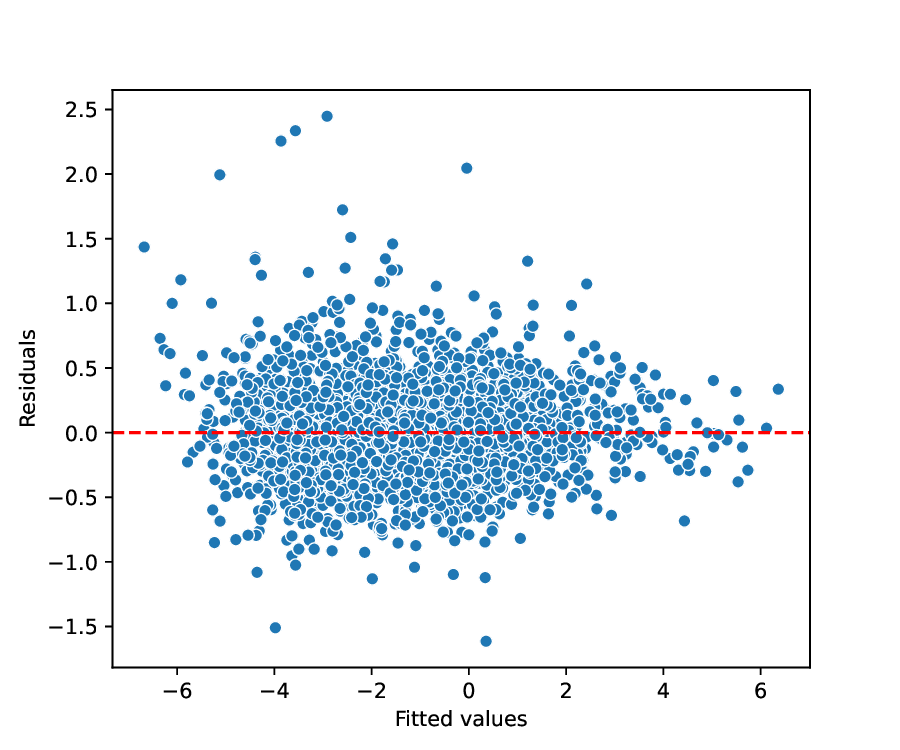}}  
\subfigure[Normal Q-Q]            {\includegraphics[width=0.49\textwidth]{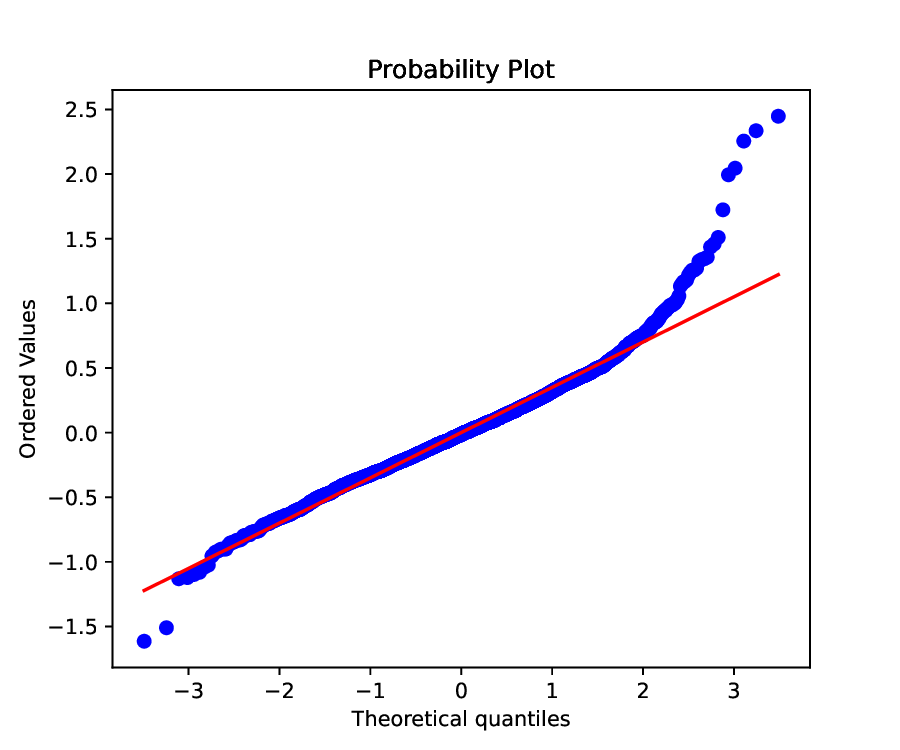}} 
\\ 
\subfigure[Scale-Location]        {\includegraphics[width=0.49\textwidth]{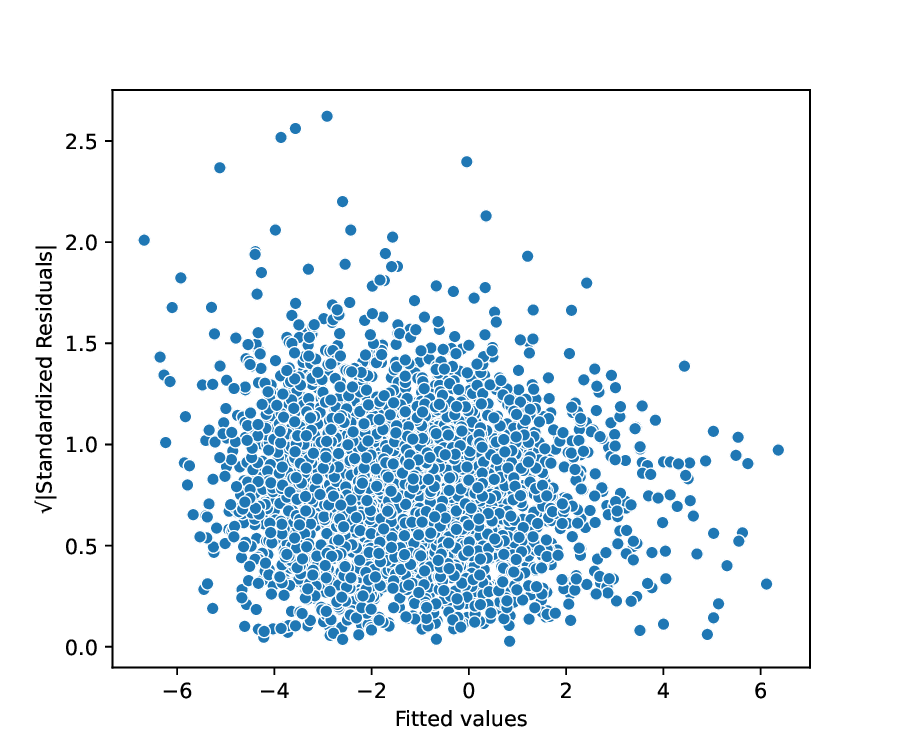}}  
\subfigure[Residuals vs Leverage] {\includegraphics[width=0.49\textwidth]{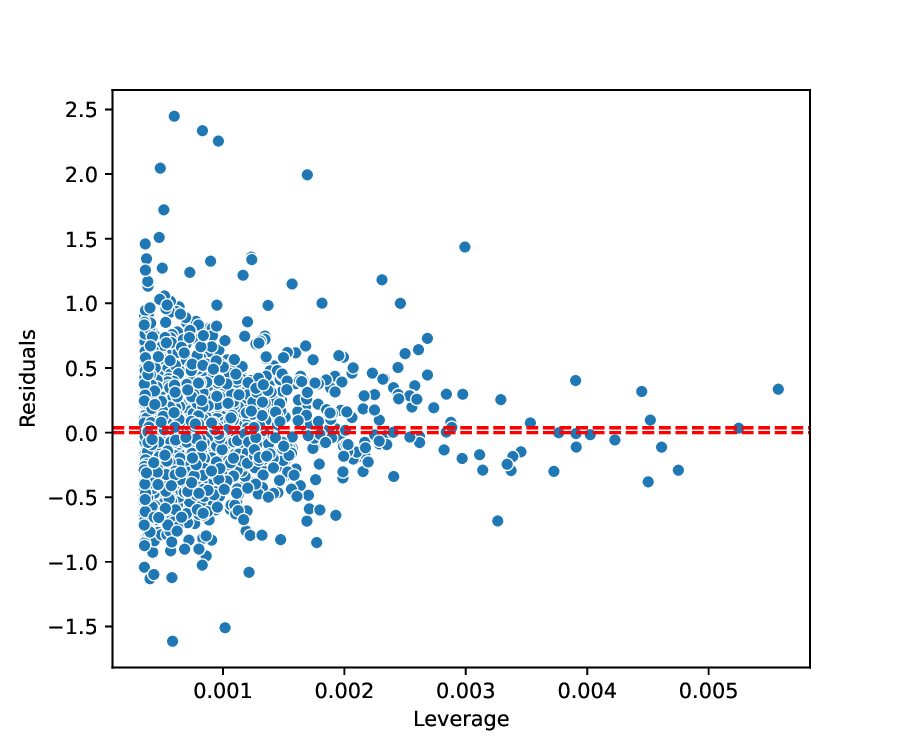}}
\caption{Diagnostic plots for model~\eqref{log CD-0}.}
\label{DP_plot-0}
\end{figure}

Taking logarithms of the two sides of equation~\eqref{CD-0}   yields
\begin{equation}\label{log CD-0}
\log\bigg({m_i^{\textrm{TI}} \over l_i}\bigg) = \alpha\log\bigg({s_i \over l_i}\bigg) + \log(\beta).
\end{equation}
The parameter $\alpha$  can be estimated using simple linear regression, which gives 
\begin{equation}\label{CI-TI}
\textrm{$\hat{\alpha}=0.990$ with the 95\% confidence interval $(0.983, 0.997)$.} 
\end{equation}
The resulting scatter-plot based on the points 
\begin{equation}\label{log CD-0sp}
\bigg( \log\bigg({s_i \over l_i}\bigg), \log\bigg({m_i^{\textrm{TI}} \over l_i}\bigg)\bigg), \quad i=1,\dots , 2,849, 
\end{equation}
and the fitted least squares regression line are depicted in Figure~\ref{CM2-0}.
The diagnostic plots are in Figure~\ref{DP_plot-0}.
Note that the 95\% confidence interval $(0.983, 0.997)$ is included in the interval $(0.5, 1)$ and therefore,   equation~\eqref{CD-0} can be assumed to hold with some $\alpha\in(0.5, 1)$.

\subsection{TF-IDF-VADER}

With the number $m_i^{\textrm{TIV}}$ of those words that are in the cleaned narrative $d_i$ as well as in the set $W_{-}$, the adjusted  dollar amount $l_i$, and the sentiment score $s_i$ defined by equation~\eqref{def-si}, the Cobb-Douglas  relationship can be written as 
\begin{equation}\label{CD}
m_i^{\textrm{TIV}} = \beta s_i^{\alpha}l_i^{1-\alpha}. 
\end{equation}
Naturally, being based on the TF-IDF-VADER method rather than TF-IDF, this equation leads to different estimates of the geometric weight $\alpha\in(0,1)$ and the scale factor $\beta>0$  than those obtained from equation~\eqref{CD-0}. As  shown in next Section~\ref{sec: input-output}, these estimates will give rise to different shapes of the input-output system's transfer functions.

\begin{note}\label{vader-4}
Since the sentiment score $s_i$ is non-negative and does not exceed $4m_i^{\textrm{TIV}}$, the statement  $s_i/(4m_i^{\textrm{TIV}}) \in [0,1]$ holds. 
\end{note}

\begin{figure}[h!]
\centering
\includegraphics[width=0.5\textwidth]{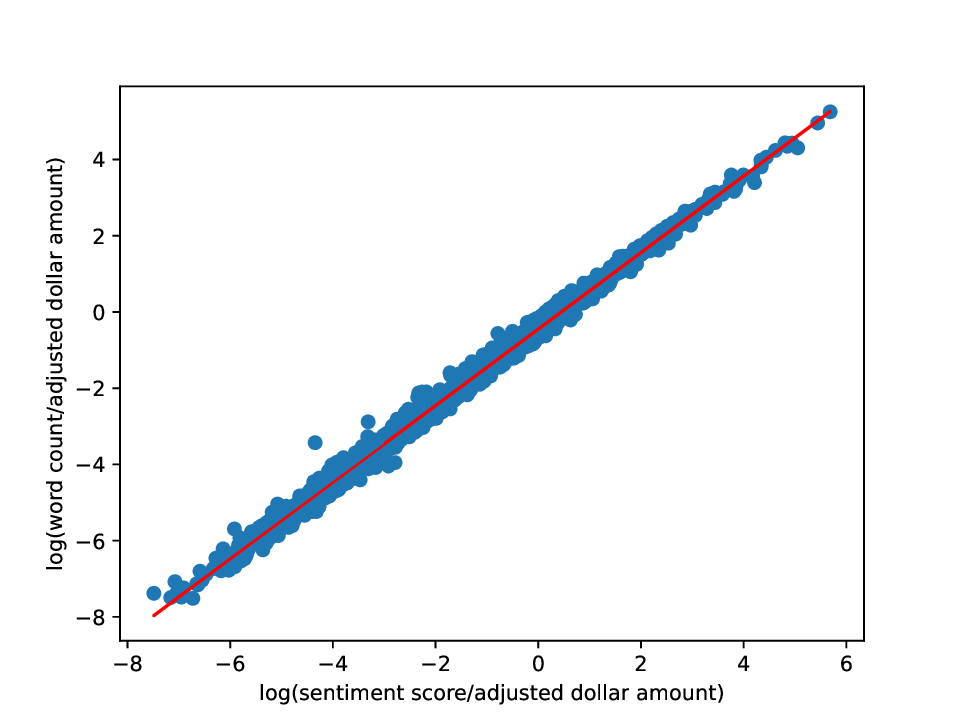}
\caption{Scatter-plot~\eqref{log CD-sp} and the fitted least squares regression line.}
\label{CM2}
\end{figure}

\begin{figure}[h!]
\centering
\subfigure[Residuals vs Fitted]   {\includegraphics[width=0.49\textwidth]{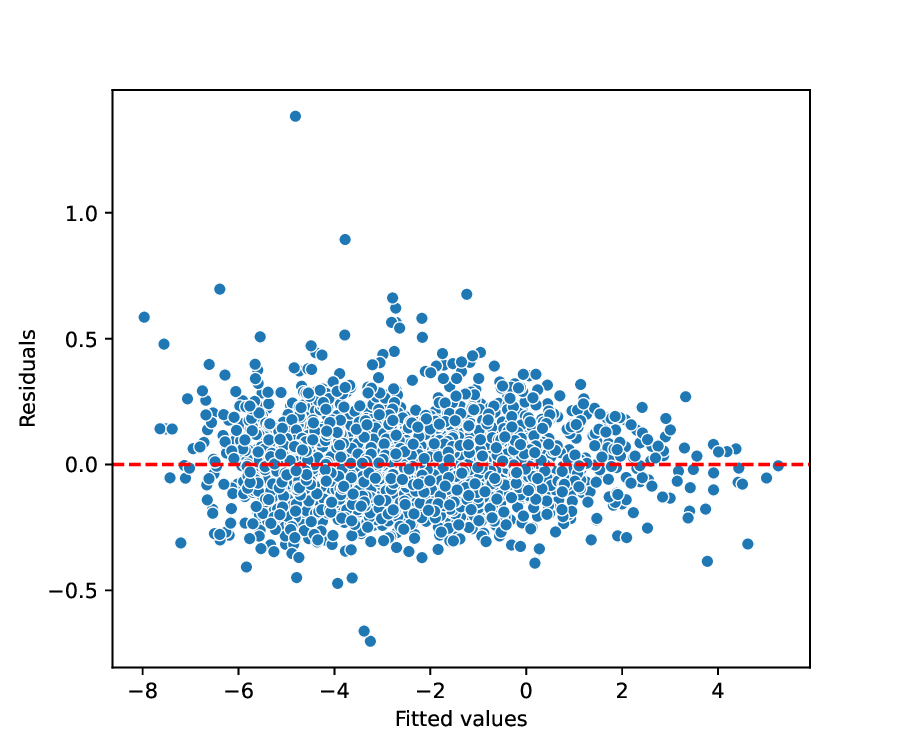}}  
\subfigure[Normal Q-Q]            {\includegraphics[width=0.49\textwidth]{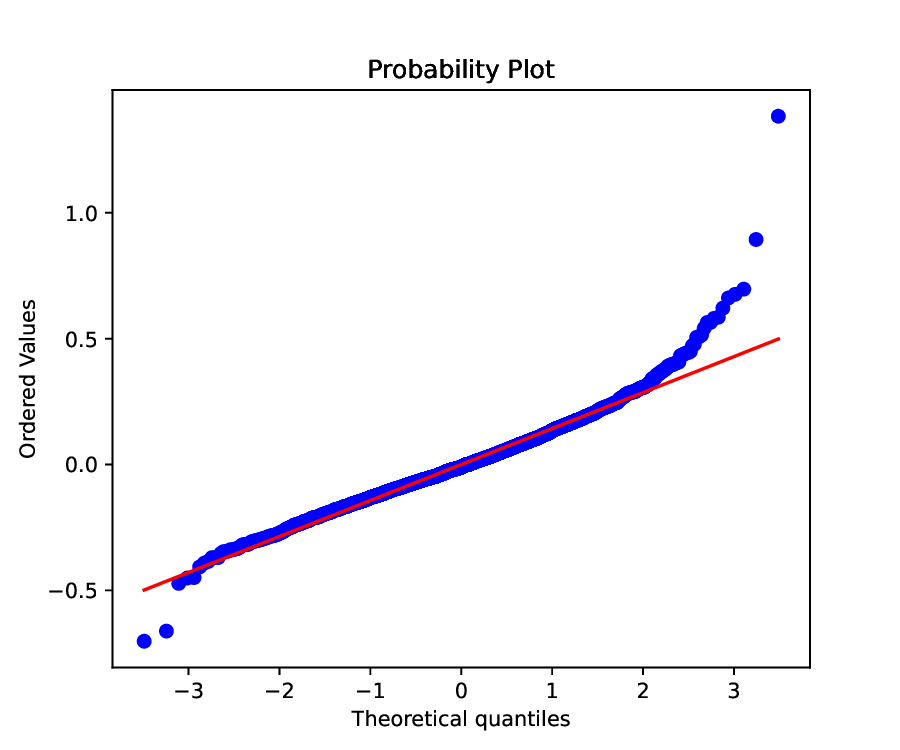}} 
\\ 
\subfigure[Scale-Location]        {\includegraphics[width=0.49\textwidth]{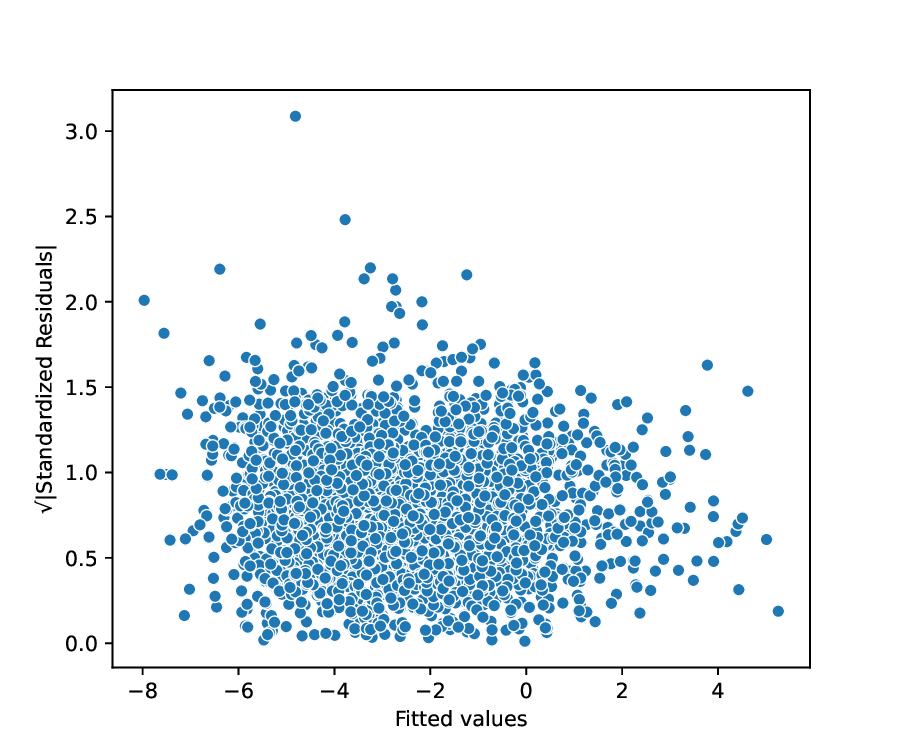}}  
\subfigure[Residuals vs Leverage] {\includegraphics[width=0.49\textwidth]{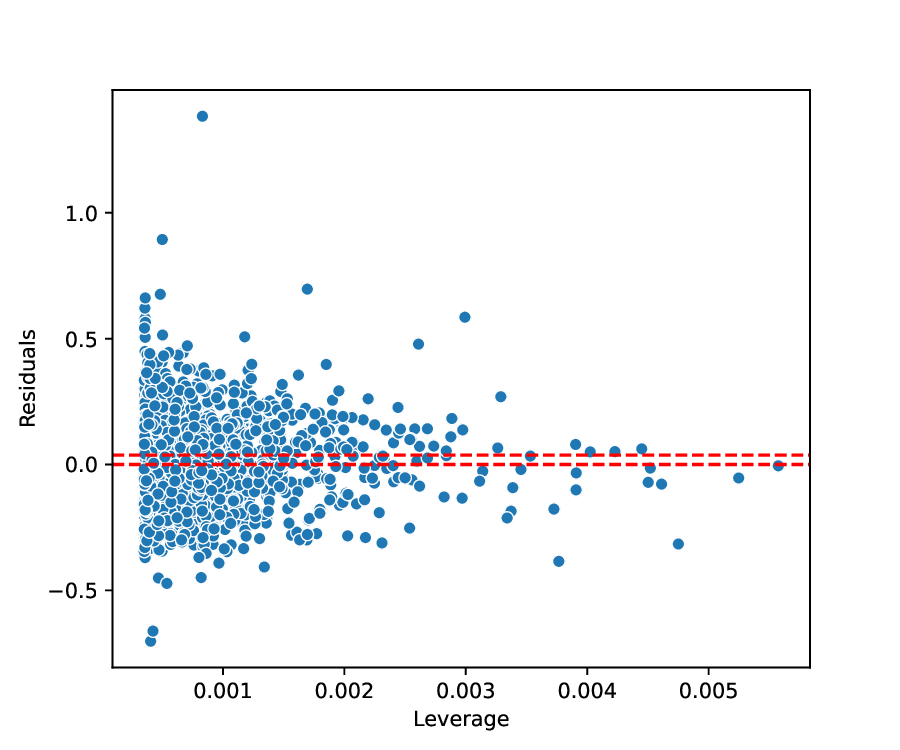}}
\caption{Diagnostic plots for model~\eqref{log CD-0}.}
\label{DP_plot}
\end{figure}

Taking logarithms of the two sides of equation~\eqref{CD}  yields 
\begin{equation}\label{log CD}
\log\bigg({m_i^{\textrm{TIV}} \over l_i}\bigg) = \alpha\log\bigg({s_i \over l_i}\bigg) + \log(\beta).
\end{equation}
The parameter $\alpha$  can be estimated using simple linear regression, which gives 
\begin{equation}\label{CI-TIV}
\textrm{$\hat{\alpha}=1.004$ with the 95\% confidence interval $(1.001, 1.007)$.} 
\end{equation}
The resulting scatter-plot based on the points 
\begin{equation}\label{log CD-sp}
\bigg( \log\bigg({s_i \over l_i}\bigg), \log\bigg({m_i^{\textrm{TIV}} \over l_i}\bigg)\bigg), \quad i=1,\dots , 2,849, 
\end{equation}
and the fitted least squares regression line are depicted in Figure~\ref{CM2}.
The diagnostic plots are in Figure~\ref{DP_plot}.
Note that the 95\% confidence interval $(1.001, 1.007)$ suggests that equation~\eqref{CD} holds for some $\alpha>1$.

\section{The input-output systems}
\label{sec: input-output}

Training anomaly detection classifiers leads to confusion matrices that involve classification errors. It also generates a meritorious subset of observations. In addition to being classified as meritorious, the observations within the meritorious set are expected to follow equations~\eqref{CD-0} or~\eqref{CD}, thus leading to an input-output system for each of the two featurizations, as we  shall describe next.

\subsection{TF-IDF}
\label{approach-ti}

First, equation~\eqref{CD-0} can be rewritten as 
$$\left(\frac{s_i}{m_i^{\textrm{TI}}}\right)^{\alpha}\left(\frac{l_i}{m_i^{\textrm{TI}}}\right)^{1-\alpha} = \frac{1}{\beta},$$
from which  the equation
\begin{equation}\label{h_0}
\log(y_i^{\textrm{TI}}) = \frac{\alpha}{1-\alpha}\log(x_i^{\textrm{TI}})+c_1 
\end{equation}
follows with the outputs 
\begin{equation}\label{outputs-y}
y_i^{\textrm{TI}} = {m_i^{\textrm{TI}} \over l_i} ,  
\end{equation}
the inputs 
\begin{equation}\label{inputs-x}
x_i^{\textrm{TI}} = {s_i \over 4m_i^{\textrm{TI}}} , 
\end{equation}
and the constant 
\[
c_1= \frac{\alpha}{1-\alpha}\log(4)+\frac{1}{1-\alpha}\log(\beta ). 
\]
Equation~\eqref{h_0} establishes the input-output relationship 
\[
y_i^{\textrm{TI}}=h_0(x_i^{\textrm{TI}})
\]
with the transfer function 
\begin{equation}\label{transfer-h0}
h_0(x)=c_2 x^{\alpha/(1-\alpha)}, 
\end{equation}
where 
\[
c_2 = e^{c_1}. 
\]
Since $x_i^{\textrm{TI}} \in [0,1]$ (recall Note~\ref{vader-4-0}), the domain of definition of $h_0$ is the unit interval $[0,1]$.

The notion of transfer function plays a pivotal role in the anomaly detection method developed by \citet{GZ2018,GZ2019a,GZ2019b,GZ2020} and \citet{syz2022}. Its importance will be seen in the next section. Specifically to function~\eqref{transfer-h0},  the first derivative of $h_0$ is 
\[
h_0'(x)=c_3 x^{(2\alpha-1)/(1-\alpha)}
\]
with a constant $c_3$. Hence, since $\alpha \in (0.5, 1)$ due to result~\eqref{CI-TI} and since the inputs are in $[0,1]$, we conclude that the derivative $h_0'$ is bounded on the interval $[0,1]$. Therefore, the transfer function $h_0$ is Lipschitz continuous on $[0,1]$. This is an important property as seen from \citet[Theorem~1~(i)]{syz2022}.

\subsection{TF-IDF-VADER}
\label{approach-tiv}

Analogously to Section~\ref{approach-ti}, the input-output relationship
\[
y_i^{\textrm{TIV}}=h_0(x_i^{\textrm{TIV}})
\]
holds with the same transfer function $h_0$ and the analogously defined inputs 
\begin{equation}\label{inputs-x-tiv}
x_i^{\textrm{TIV}} = {s_i \over 4m_i^{\textrm{TIV}}} 
\end{equation}
and outputs 
\begin{equation}\label{outputs-y-tim}
y_i^{\textrm{TIV}} = {m_i^{\textrm{TIV}} \over l_i} .   
\end{equation}
However, in view of result~\eqref{CI-TIV}, the assumption $\alpha>1$   has to be considered, in which case the transfer function 
\[
h_0(x)=c_2 x^{\alpha/(1-\alpha)} 
~\bigg( = { c_2\over x^{\alpha/(\alpha-1)}} \bigg) 
\]
is not Lipschitz continuous on the interval $[0,1]$. Fortunately, it satisfies another set of conditions of \citet[Theorem~1~(ii)]{syz2022} enabling us to apply the anomaly detection method.

\section{Anomaly detection among quantified narratives}
\label{sec: performance}

In the ideal scenario, for the $i$'s in the predicted meritorious set, the pairs $(x_i, y_i)$,  which are either $(x_i^{\textrm{TI}},y_i^{\textrm{TI}})$ or $(x_i^{\textrm{TIV}},y_i^{\textrm{TIV}})$ depending on the featurization method, form an anomaly-free input-output system. In reality, anomalies are inevitable for any input-output system that is not particularly stringent. Anomalies may even be present in data due to the very nature of the problem. However, they may not always drastically change the regime of inputs and/or outputs, and may therefore be difficult to detect for typical classifiers. 

For this reason, two indices   are next provided based on the pairs $(x_i, y_i)$ within the predicted meritorious set.  The indices are constructed by adapting the method for systematic-anomaly detection developed by \citet{GZ2018,GZ2019a,GZ2019b,GZ2020} in the case of (quantitative) i.i.d. inputs and subsequently extended to time series inputs by \citet{syz2022}.

Generally speaking, the indices help to see if there is any background or systematic risk in the input-output system characterized by transfer functions $h_0$. Namely, they help to distinguish the ``null hypothesis'' $y = h_0(x)$ from the ``alternative'' $y = h_0(x + \delta)$, where the unspecified random quantity $\delta$ represents the exogenous background risk that naturally arises in many problems \citep{GZ2018}. An advantage of these systematic-anomaly detection indices is that neither the transfer function $h_0$ nor the distribution of $\delta$ need to be specified. 
 
In the case of this paper, the featurized text inputs may or may not contain a background risk, and the transfer function $h_0$ is based on some estimated parameters, whose actual values are not known. Hence, detecting the background risk when it exists can be done by checking the asymptotic behavior of the indices without specifying the background risk or the transfer function.  This feature is especially helpful in exploratory studies such as the present one. Of course, certain conditions are needed for the development of a rigorous theory, but this is outside the scope of the present paper. For technical details, we refer to \citet{syz2022} and references therein.

\subsection{The $I$-index and findings}

Let a simple random sample of pairs $(x_1, y_1), \dots, (x_n, y_n)$ form an input-output system, where the inputs $x_i$ and the outputs $y_i$ are assumed to be realizations of continuous random variables. Suppose that there are no ties among $x_1,\ldots,x_n$, in which case the inputs  can be ordered in a unique way,   denoted as $x_{1:n} < x_{2:n} < \cdots < x_{n:n}$.   The corresponding concomitants \citep[see, e.g.,][]{DN2003} of the outputs $y_1,\ldots,y_n$  are denoted by $y_{1,n}, \dots, y_{n,n}$.

Relying on the theory developed by \citet{GZ2018,GZ2019a,GZ2019b,GZ2020} and \citet{syz2022},  the presence of systematic anomalies among the observed pairs can be assessed by the rate of convergence to $1/2$ of the $I$-index 
\[
I_{n} = {\sum^n_{i=2} \big( y_{i,n} - y_{i-1,n}\big)_{+} \over \sum^n_{i=2}\big| y_{i,n} - y_{i-1,n}\big| }, 
\]
where $a_{+}=a$ when $a> 0$ and $a_{+}=0$ when $a\le 0$.

The $I$-index appeared in \citet{dz2017} as the solution to an optimization problem. It was further explored by \citet{cdgz2018} and subsequently extended to the multivariate case by \citet{dmz2019} to accommodate  diverse  applications. The index has found uses in, e.g., education sciences \citep{cz2017}, but most notably in various branches of engineering and related areas, such as system security \citep{GZ2019a,GZ2020}, anomaly detection \citep{syz2022}, electrical engineering \citep{Chernov(2021),Di Capua(2023a),Di Capua(2023b),Milano(2023)}. mechanical and materials engineering \citep{Kirk(2020),kma2021,Sauceda(2022)},
networks \citep{Gurfinkel(2020),GR(2020)}, and prognostics \citep{Trilla(2020)}.

\begin{figure}[h!]
\centering
\subfigure[LR]  {\includegraphics[width=0.49\textwidth]{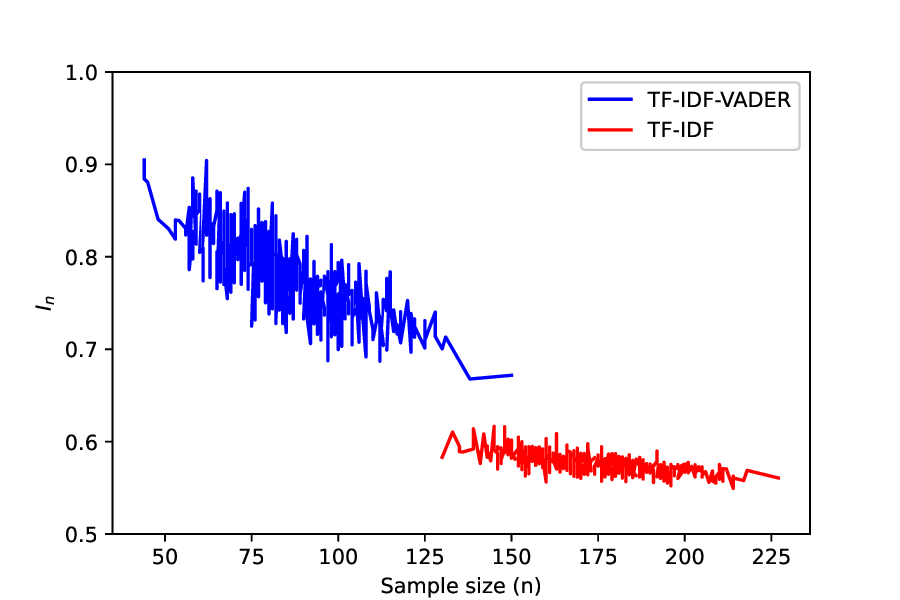}} 
\subfigure[SVM] {\includegraphics[width=0.49\textwidth]{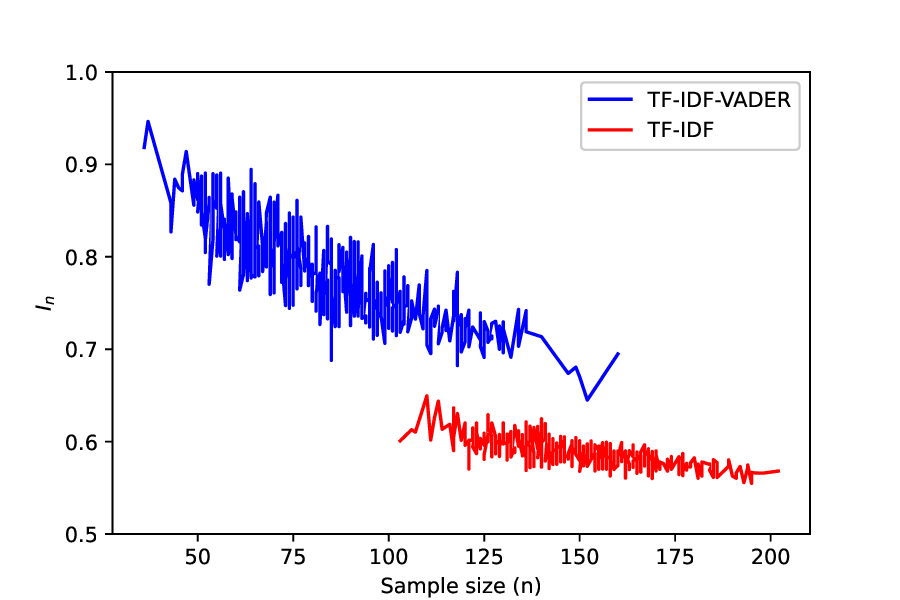}}
\\
\subfigure[GB]  {\includegraphics[width=0.49\textwidth]{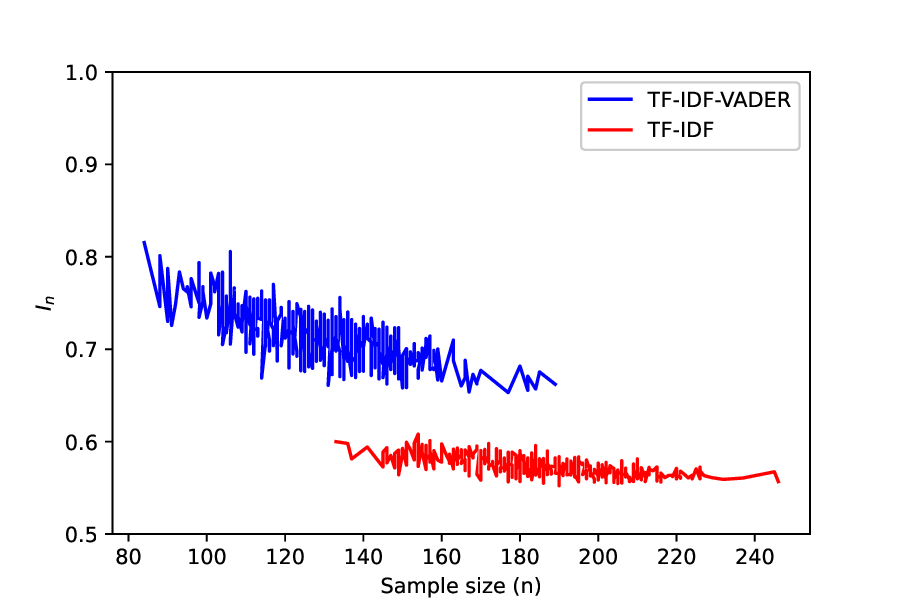}} 
\subfigure[MLP] {\includegraphics[width=0.49\textwidth]{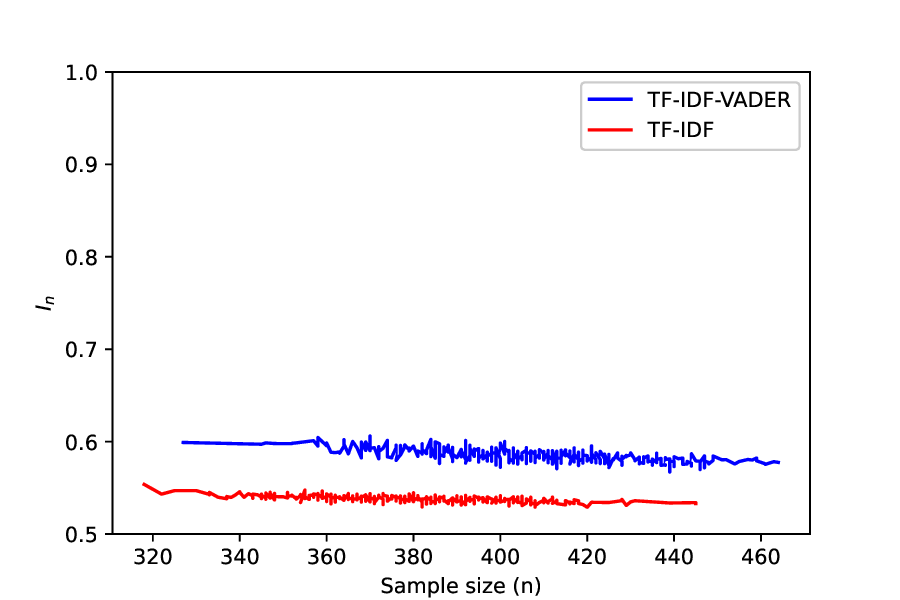}}
\\
\subfigure[RF]  {\includegraphics[width=0.49\textwidth]{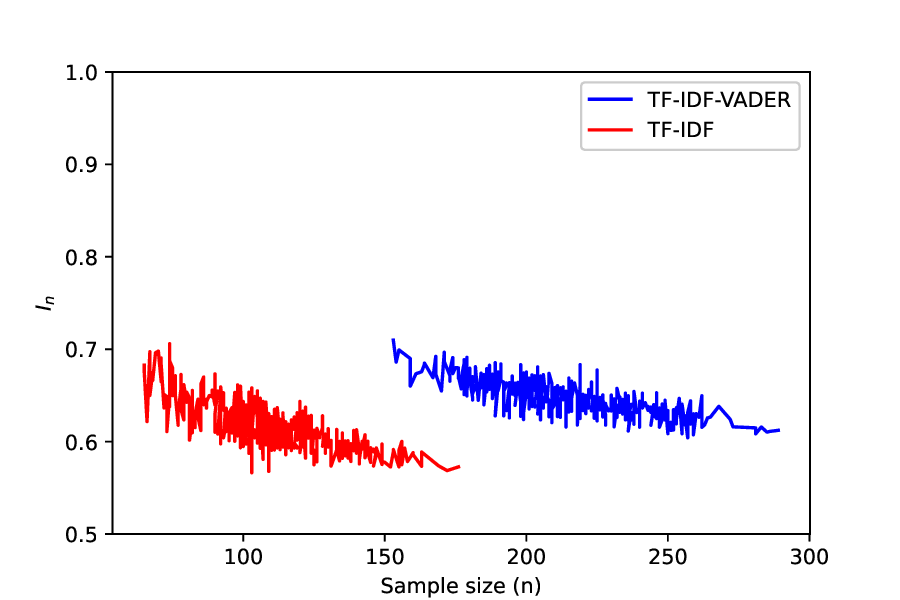}}
\caption{The plots of $I_n$ for the five classifications.}
\label{Ivalues}
\end{figure}

Applying Algorithm~\ref{algorithm-0}  
\begin{table}[h!]  
\centering
\begin{tabular}{l}
\hline 
{\bf Require} 
\\
{1.} A dataset $D$ with labels ``meritorious'' and ``non-meritorious''. \\ 
{2.} A specific  classification method. 
\\ \hline 
{\bf Compute a value of the $I$-index}   
\\
{1.} Randomly assign $(1-r)100\%$ of the dataset $D$ as a training set, and  
the remaining \\ \quad $r100\%$ as a testing set.
\\
{2.} Train a classifier $f$ using the specific classification method based on the training set.
\\
{3.} Apply $f$ on the testing set, leading to a meritorious subset with the indices  
$d_m\subseteq D$ of \\ \quad the testing set.
\\
{4.} Compute the $I$-index  using $(x_i,y_i)$, $i\in d_m$.
\\ \hline 
\end{tabular}
\renewcommand{\tablename}{Algorithm} 
\caption{Calculating the $I$-index.}
\label{algorithm-0}
\end{table}
on each of the predicted meritorious sets generated by the five classifiers, we obtain  
Figure~\ref{Ivalues} 
depicting the $I$-index for each of the five classifiers. In Algorithm~\ref{algorithm-0}, the parameter $r$ is set to $0.4$.

Note that for these classifiers except for the MLP, the graphs rapidly move in the direction of $1/2$, although we cannot say with certainty that they will reach $1/2$ if the sample size tends to infinity. At any rate,  given what we see, it can  be concluded that systematic anomalies are present. As to the MLP classifier, it is difficult to claim from the nearly flat graphs in the case of both featurizations that there is a tendency to convergence to  $1/2$, although it could be so but at a very slow rate. If it does not converge to $1/2$, then the properties of the $I$-index imply that the input-output system is anomaly free. 

Given the aforementioned ambiguities, which are natural in any exploratory study such as the present one, it is prudent to have a supplementary view.  This leads to the next section concerning the $B$-index.

\subsection{The $B$-index and findings}

\begin{figure}[h!]
\centering
\subfigure[LR]  {\includegraphics[width=0.49\textwidth]{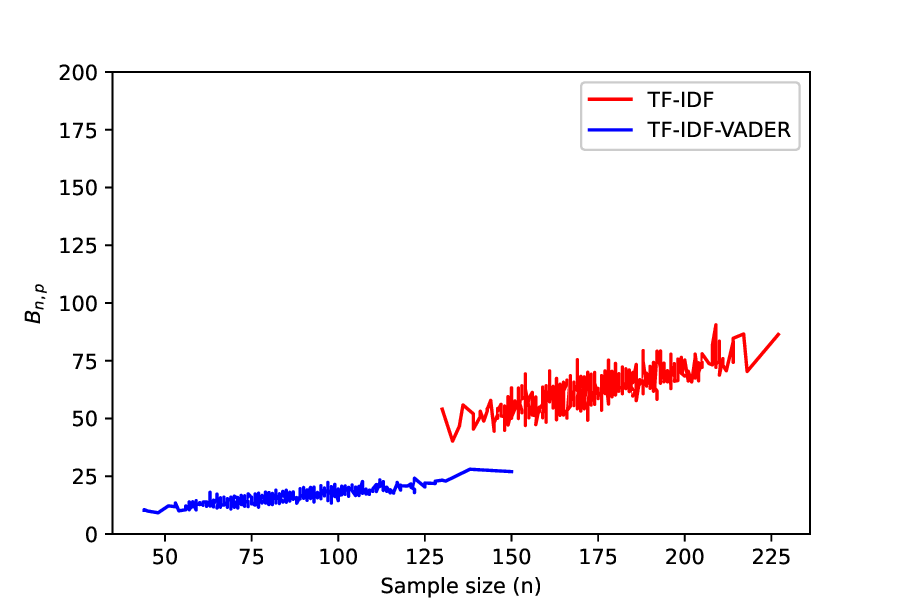}} 
\subfigure[SVM] {\includegraphics[width=0.49\textwidth]{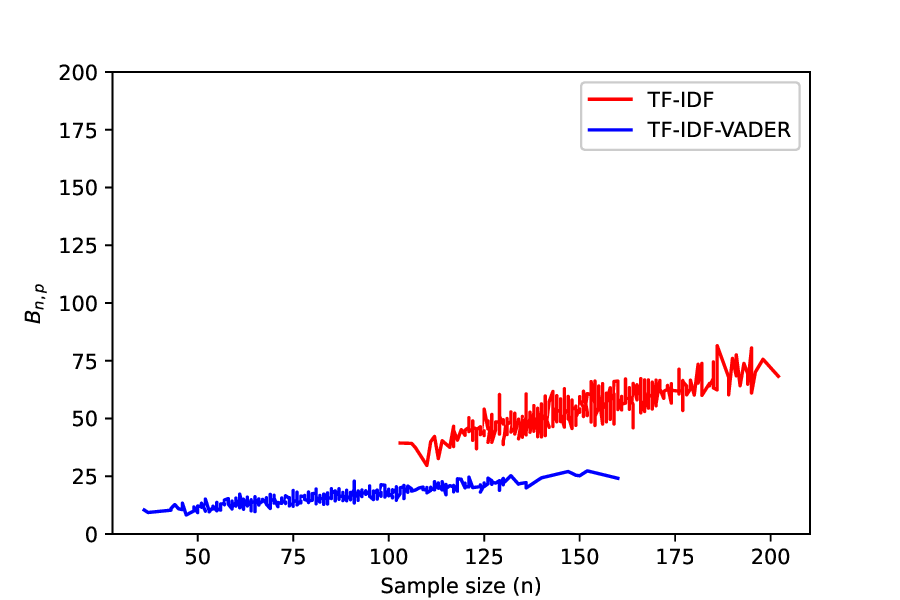}}
\\
\subfigure[GB]  {\includegraphics[width=0.49\textwidth]{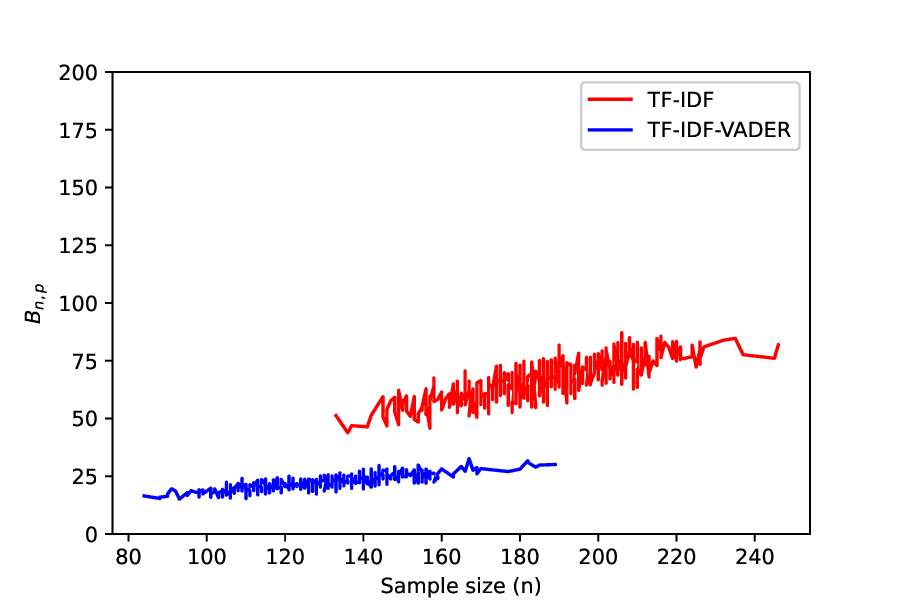}} 
\subfigure[MLP] {\includegraphics[width=0.49\textwidth]{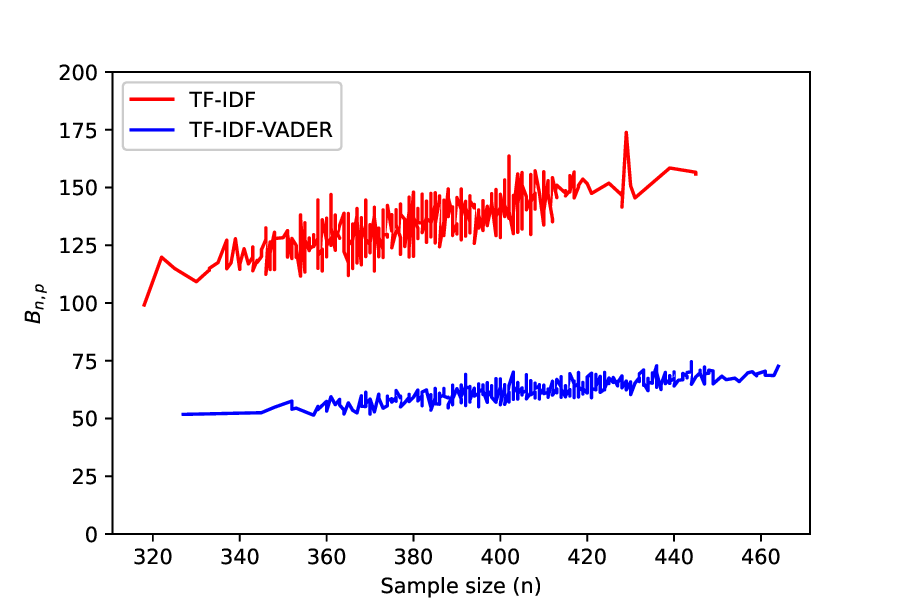}}
\\
\subfigure[RF]  {\includegraphics[width=0.49\textwidth]{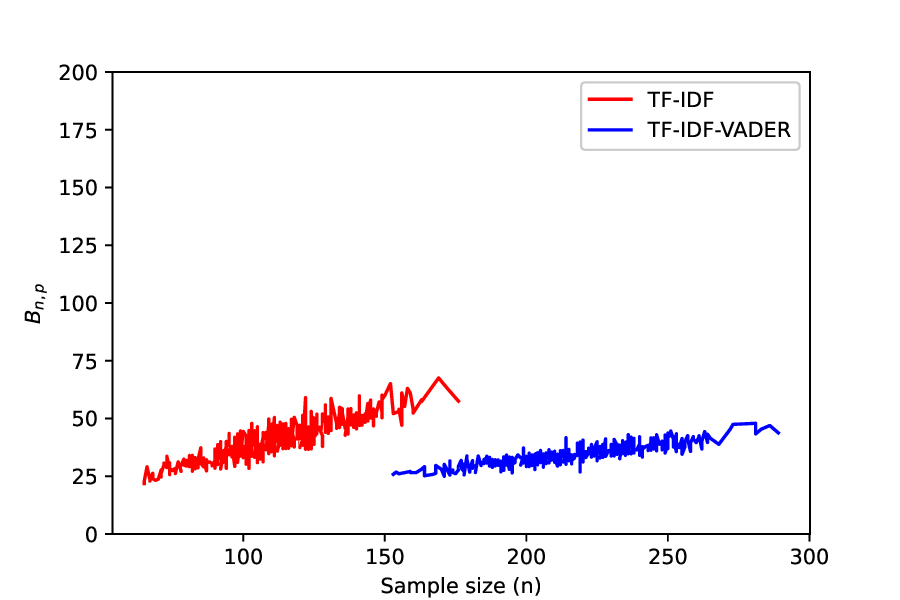}}
\caption{The plots of $S_n$ for the five classifications.}
\label{B when p=0}
\end{figure}

To introduce the $B$-index,  note that for any real number $a$ its positive part $a_{+}$  can be expressed as $(a+|a|)/2$. Since $\sum^n_{i=2} \big( y_{i,n} - y_{i-1,n}\big)=  y_{n,n}- y_{1,n}$,  it follows that 
\begin{align*}
I_{n} 
&= {\sum^n_{i=2} \big( y_{i,n} - y_{i-1,n}\big)_{+} \over \sum^n_{i=2}\big| y_{i,n} - y_{i-1,n}\big| }
\\
&= {1\over 2} \bigg ( 1+  { y_{n,n}- y_{1,n} \over \sum^n_{i=2}\big| y_{i,n} - y_{i-1,n}\big| } \bigg).  
\end{align*}
In view of this representation of the $I$-index, the convergence of $I_n$ to $1/2$ is associated with the stochastic growth to infinity of the sum 
\[
S_n= \sum^n_{i=2}\big| y_{i,n} - y_{i-1,n}\big| 
\] 
when $n\to \infty $.  This sum  is depicted as a function of $n$ in Figure~\ref{B when p=0} 
for the five classifications. To produce the figure, Algorithm~\ref{algorithm-0}   has been implemented with $S_n$ instead of $I_n$, setting the parameter $r$ to $0.4$. For the purpose of comparison, the vertical axes in all the five panels of the figure are the same and run from $0$ to $200$.

To parameterize the stochastic growth of the sum $S_n$ to infinity when $n$ increases, the $B$-index 
\[
B_{n,p} = {1\over n^{1/p}} \sum^n_{i=2}\big| y_{i,n} - y_{i-1,n}\big| 
~\bigg ( = {S_n\over n^{1/p}} \bigg) 
\]
is considered, where $p\in (0,\infty ]$ is a parameter. The normalization $n^{1/p}$ mimics that of the law of large numbers when $p=1$ and that of the central limit theorem when $p=2$. In the extreme case $p=\infty $, the normalization is identically equal to $1$, and thus the $B$-index reduces to $S_n$.

By varying the parameter $p$ values, the growth to infinity of $S_n$   can be sensed when $n\to\infty$. This is linked to the notion of {\it reasonable order.} Specifically, the outputs are said to be {\it in $p$-reasonable order} with respect to the inputs 
for some $p$ if $B_{n,p} $ is stochastically bounded when $n\to\infty$. Otherwise, that is, when  $B_{n,p} $ stochastically grows to infinity when  $n\to\infty$, the outputs are said to be {\it out of $p$-reasonable order}. This notion of in/out of $p$-reasonable order in the case $p=2$ was introduced by \citet{GZ2018,GZ2019a,GZ2019b,GZ2020} and subsequently extended to arbitrary $p\in (0,\infty ]$ by \citet{syz2022} to accommodate a whole spectrum of tail heaviness in time series data. 

By \citet[Theorem~4]{syz2022}, the anomaly-free outputs are in $p$-reasonable order for $p=\infty $ and therefore for all $p\in (0,\infty]$. For any two parameter values $p_1 < p_2$, the outputs, which may contain anomalies, can be out of $p_2$-reasonable order but in $p_1$-reasonable order. Hence, by visualizing the behaviour of $B_{n,p_1}$ and $B_{n,p_2}$ with respect to various sample sizes $n$, we can understand when, with respect to $p$, the meritorious set becomes in or out of $p$-reasonable order. For this task, Algorithm~\ref{algorithm-0}  is implemented with $B_{n,p}$ instead of $I_n$. We do so $k=500$ times and obtain $k$ values of the $B$-index, always setting the parameter $r$ to $0.4$.

Figures~\ref{fig:Bnp LR}--\ref{fig:Bnp RF}, which are relegated to Appendix~\ref{B-figures} due to space considerations, depict the $B$-indices arising from the TF-IDF and TF-IDF-VADER featurizations.  For the purpose of comparison, the vertical axes in all the panels of the figures are the same and run from $0$ to $5$.  It is observed that in the case of both featurizations and for every classification method, when the parameter $p$ decreases, that is, when its reciprocal $1/p$ goes through the values $0.6$, $0.7$, $0.8$ and $0.9$, the meritorious sets shift at some point from being out of $p$-reasonable order to being in $p$-reasonable order. This pattern is, of course, natural as the normalizing factor $n^{1/p}$ gets heavier for larger $1/p$, thus slackening the growth of the sum $S_n$ when systematic anomalies are present.

From Figures~\ref{fig:Bnp LR}--\ref{fig:Bnp GB}, which refer to the LR, SVM, and GB classifiers, the TF-IDF featurization generally leads to larger sample sizes than the TF-IDF-VADER featurization. In the MLP case,  it can be seen from Figure~\ref{fig:Bnp MLP} that the sample sizes, in general, become similar. In the RF case, as seen from Figure~\ref{fig:Bnp RF}, it is the TF-IDF-VADER featurization that generally leads to larger sample sizes than the TF-IDF featurization.

Note also from Figures~\ref{fig:Bnp LR}--\ref{fig:Bnp RF} that in the case of the TF-IDF featurization, the system tends to shift from  out of $0.7^{-1}$-reasonable order to in $0.8^{-1}$-reasonable order, whereas in the case of the TF-IDF-VADER featurization, the tendency to shift from out of $0.6^{-1}$-reasonable order to in $0.7^{-1}$-reasonable order  is observed.

To formally check if the resulting meritorious set is in $p$-reasonable order,  an index
$p^*$  is defined as the largest $p\ge 1$ such that the $B$-index  $B_{n,p}$ is stochastically bounded when $n\to \infty $. Of course, depending on the featurization, there are two $p^*$'s: one is $p^*_{\text{TI}}$ arising from the TF-IDF featurization, and the second one is  $p^*_{\text{TIV}}$ arising from the TF-IDF-VADER featurization.  It is seen from the figures that, roughly speaking,  
\[
p^*_{\text{TI}}\in(0.8^{-1}, 0.7^{-1}) \quad \textrm{and} \quad  
p^*_{\text{TIV}}\in(0.7^{-1}, 0.6^{-1})
\]
for all the five classification methods.  
This implies $p^*_{\text{TIV}} > p^*_{\text{TI}}$ regardless of the classification method and suggests that there exists $p'\in(p^*_{\text{TI}}, p^*_{\text{TIV}})$ such that, for any given classification method, the $B$-index $B_{n,p'} $ is stochastically bounded if the TF-IDF-VADER featurization is used but stochastically grows to infinity if the TF-IDF featurization is used. Therefore, the TF-IDF-VADER  featurization leads to a relatively higher extent of anomaly absence because the $B$-index  $B_{n,p'} $ rejects the TF-IDF case as being anomaly free but cannot reject the TF-IDF-VADER case as being such. When making such interpretations, smaller $p^*$'s   always suggest heavier anomalies in the system as they give rise to larger $(1/p)$'s ($p < p^*$), thus producing heavier normalizing factors $n^{1/p}$ needed to slacken the growth of the sum $S_n$ with respect to $n\to \infty$.

\section{Conclusion}
\label{sec: conclusion}

A method for detecting systematic anomalies among texts such as consumer complaint narratives  has been proposed. In addition to classification algorithms, quantitative indices have been employed to assists in anomaly detection. The indices are essential for detecting those anomalies that mimic meritorious complaints without changing distributional patterns of the word counts, discounted dollar amounts, and sentiment scores. They also play pivotal roles when automating the procedure. Repeated classifications on randomly segmented training and testing datasets have been employed to visualise the index performance in different scenarios. 

All the procedural steps have been illustrated using consumer complaints from the Consumer Complaint Database of the Consumer Financial Protection Bureau.  Our results have shown that the TF-IDF-VADER featurization outperforms the TF-IDF featurization in terms of the $I$- and $B$-indices regardless of the chosen classification method, although the TF-IDF featurization consistently leads to slightly higher classification accuracy than the TF-IDF-VADER featurization. This suggests that, in practice, the TF-IDF-VADER featurization could be applied to consumer complaints to identify those with higher priority to receive reliefs, because the resulting meritorious group is less likely to contain non-meritorious complaints.

\section*{Acknowledgements}

We are indebted to two anonymous referees for queries and suggestions that helped us to clarify and improve the manuscript. 
This research is a part of the project ``The State of the Art in Anomaly Detection and Model Construction, with the Focus on Natural Language Processing (NLP) in Actuarial Modelling'' supported by the Committee on Knowledge Extension Research (CKER) of the Society of Actuaries (SOA) Research Institute, and the Casualty Actuarial Society (CAS). The research has also been partially supported by the NSERC Alliance--MITACS Accelerate grant entitled ``New Order of Risk Management: Theory and Applications in the Era of Systemic Risk'' from the Natural Sciences and Engineering Research Council (NSERC) of Canada, and the national research organization Mathematics of Information Technology and Complex Systems (MITACS) of Canada, as well as by the individual NSERC Discovery Grant ``Automated Statistical Techniques for Systematic Anomaly Detection in High Frequency Data'' of R.~Zitikis.

\section*{Declarations of interest}

The authors report no conflicts of interest. The authors alone are responsible for the content and writing of the paper.

\bibliographystyle{apalike}

\clearpage 
\newpage 
\appendix 
\section{Steps for corpus cleaning}

The collection of those complaint narratives that we analyze is referred to as the ``corpus''. As suggested by \citet{vgpz2023}, the following steps  have been implemented to clean the narratives in the corpus:
\begin{enumerate}[(1)]
    \item Ignoring cases, i.e., converting each uppercase letter into the corresponding lowercase one.
    \item Removing punctuations, except semantic punctuations such as ``!" and ``?".
    \item Removing stop words (common words with no sentiment information) as well as dollar amounts from the corpus.
    \item Removing common and frequent words (e.g., frequent nouns that do not have sentiment information) from the corpus.
    \item  Removing the sentences that begin with ``thanks" or ``thank you''. 
\end{enumerate}
Table~\ref{text clean} illustrates these steps using excerpts from several complaint narratives accompanied with their cleaned versions. 
\begin{table}[h!]
\centering 
\begin{tabular}{p{.45\textwidth}|p{.45\textwidth}}
\hline\hline
\textbf{Original narratives} & \textbf{Cleaned narratives} \\
\hline
Macys did not reverse out my \$230.00  dispute Similar to cfpg XXXX  & macys not reverse dispute similar cfpg \\
\hline
Inquired about \$300.00 increase and did not receive a response to the request & inquired increase not receive response request \\
\hline
Order cancelled and never delivered. still charged full amount of \$7000.00 on my credit card & order cancelled never delivered still charged full amount credit card \\
\hline
A charge of \$170.00 was made to my card I did not make or authorize charge. & charge made card not make authorize charge  \\
\hline
Someone fraudulently charged \$750.00 from XXXX on our card and Citi will not take it off!!!!!!!!!! & someone fraudulently charged card citi not take !!!!!!!!!!\\
\hline
\end{tabular}
\caption{Selected complaint narratives and their cleaned versions.}
\label{text clean}
\end{table}

When cleaning, the dollar amounts  are removed because the VADER approach is unable to ascertain the sentiment scores associated with numerical values.  However, the exclamation ``!'' and question ``?'' marks  are not removed because they have strong   VADER sentiment intensities. That is, different VADER sentiment intensities can arise for the same narratives with or without these  two punctuation marks.

\section{The $B$-index and samples sizes}
\label{B-figures}

\begin{figure}[h!]
\centering
\subfigure[$p=0.6^{-1}$] {\includegraphics[width=0.49\textwidth]{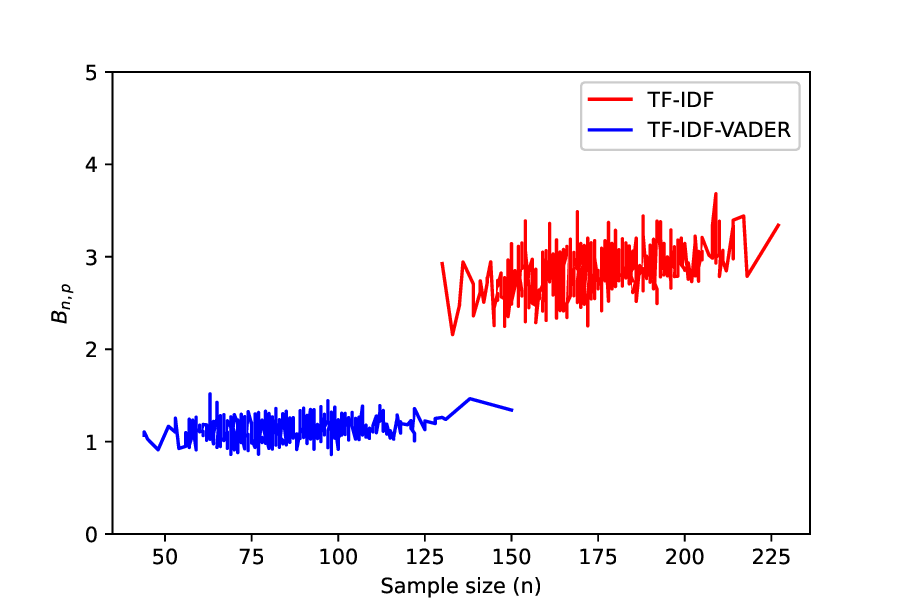}}  
\subfigure[$p=0.7^{-1}$] {\includegraphics[width=0.49\textwidth]{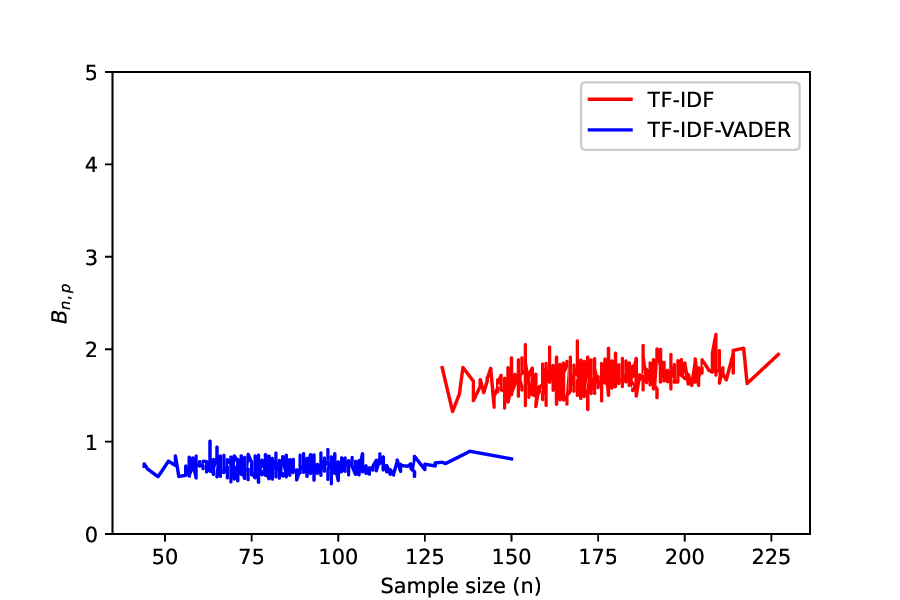}} 
\\ 
\subfigure[$p=0.8^{-1}$] {\includegraphics[width=0.49\textwidth]{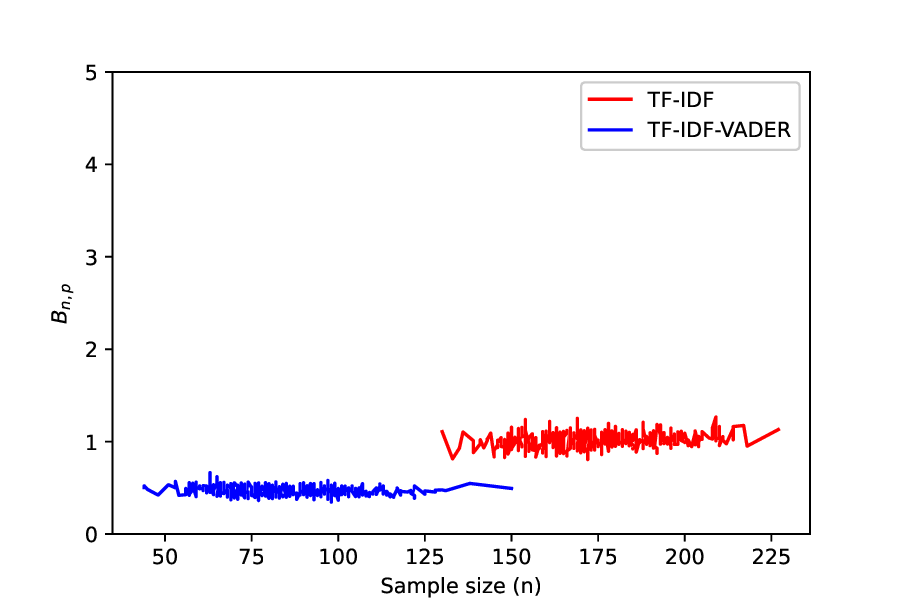}}  
\subfigure[$p=0.9^{-1}$] {\includegraphics[width=0.49\textwidth]{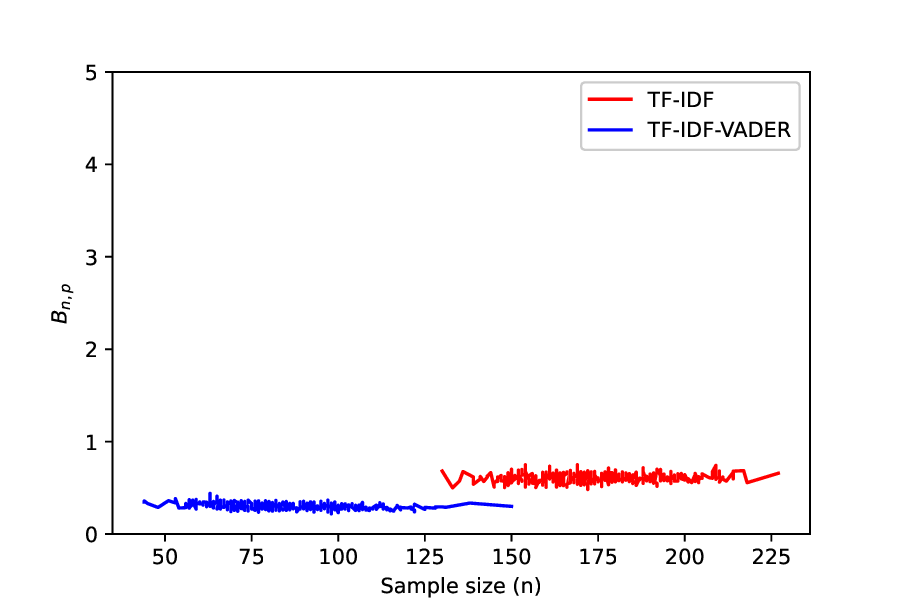}}
\caption{The plots of $B_{n,p}$ (vertical axis) for various sample sizes $n$ (horizontal axis) and parameter $p$ values when using the LR classification.}
\label{fig:Bnp LR}
\end{figure}

\begin{figure}[h!]
\centering
\subfigure[$p=0.6^{-1}$] {\includegraphics[width=0.49\textwidth]{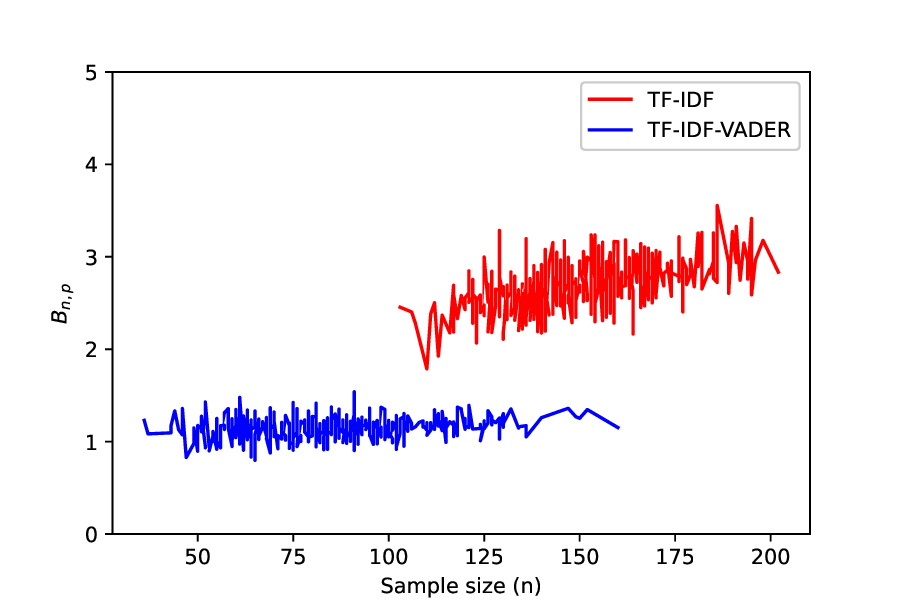}}  
\subfigure[$p=0.7^{-1}$] {\includegraphics[width=0.49\textwidth]{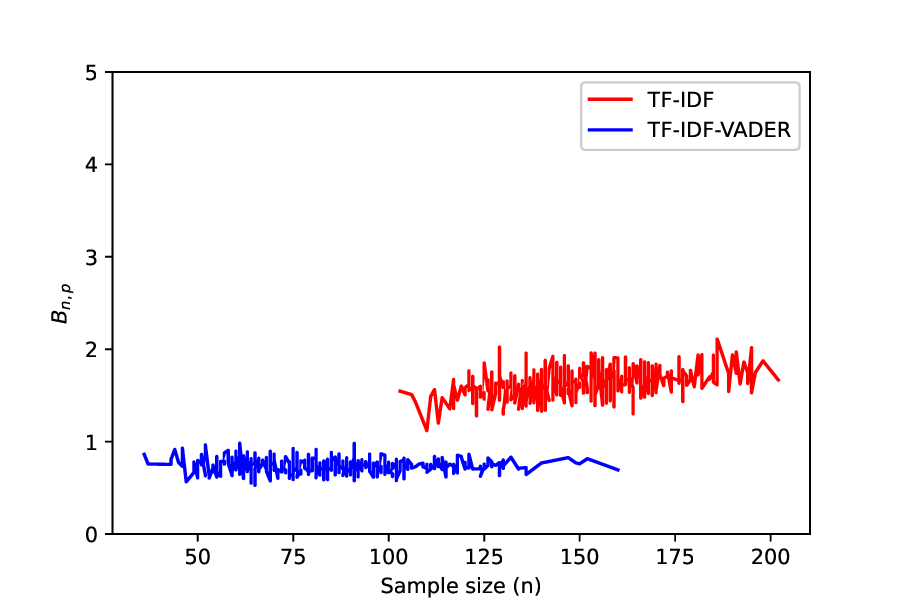}}
\\
\subfigure[$p=0.8^{-1}$] {\includegraphics[width=0.49\textwidth]{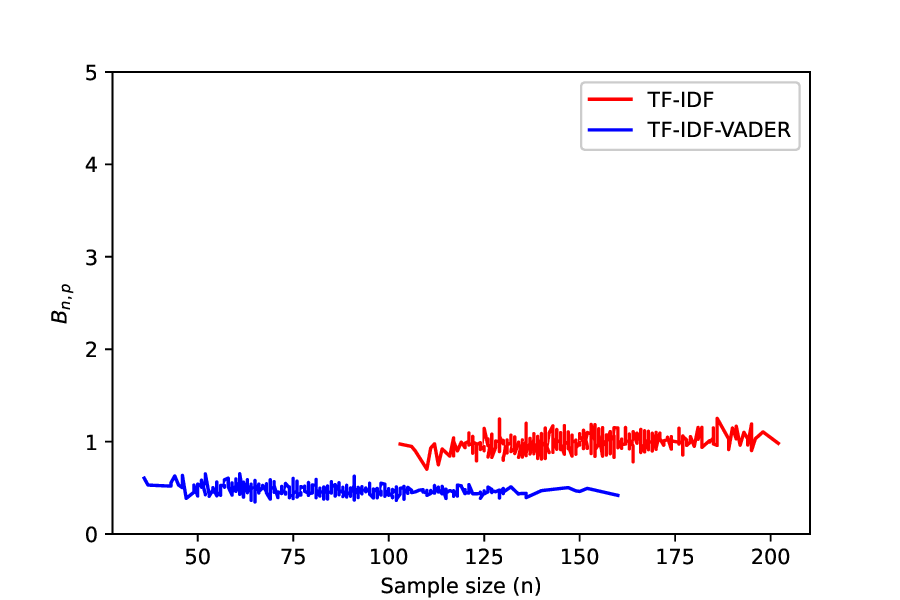}}  
\subfigure[$p=0.9^{-1}$] {\includegraphics[width=0.49\textwidth]{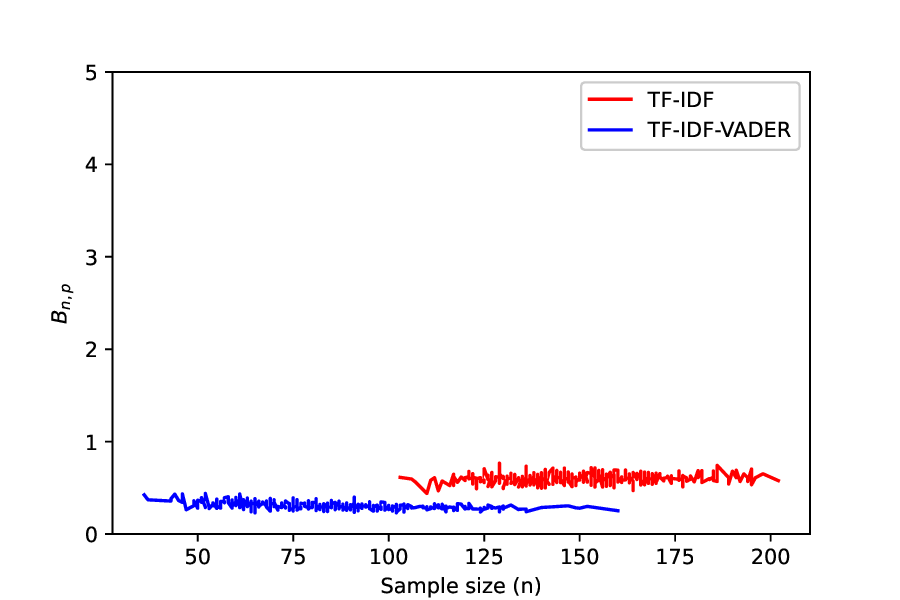}}
\caption{The plots of $B_{n,p}$ (vertical axis) for various sample sizes $n$ (horizontal axis) and parameter $p$ values when using the SVM classification.}
\label{fig:Bnp SVM}
\end{figure}
 
\begin{figure}[h!]
\centering
\subfigure[$p=0.6^{-1}$] {\includegraphics[width=0.49\textwidth]{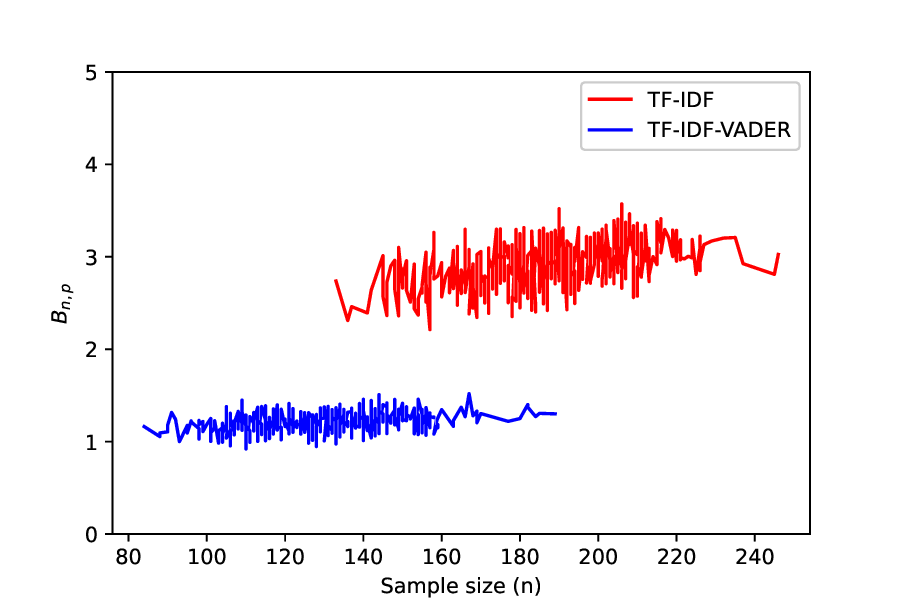}}  
\subfigure[$p=0.7^{-1}$] {\includegraphics[width=0.49\textwidth]{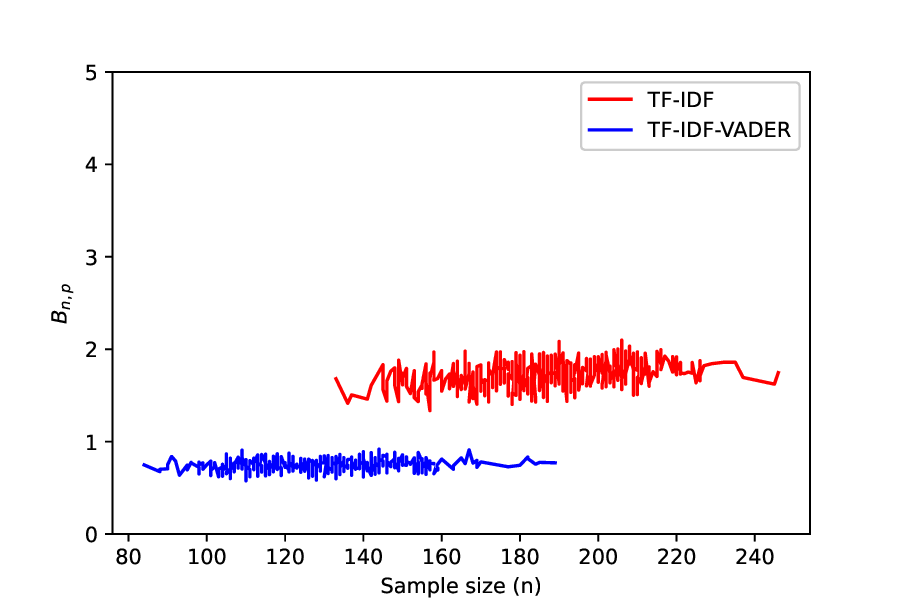}}
\\
\subfigure[$p=0.8^{-1}$] {\includegraphics[width=0.49\textwidth]{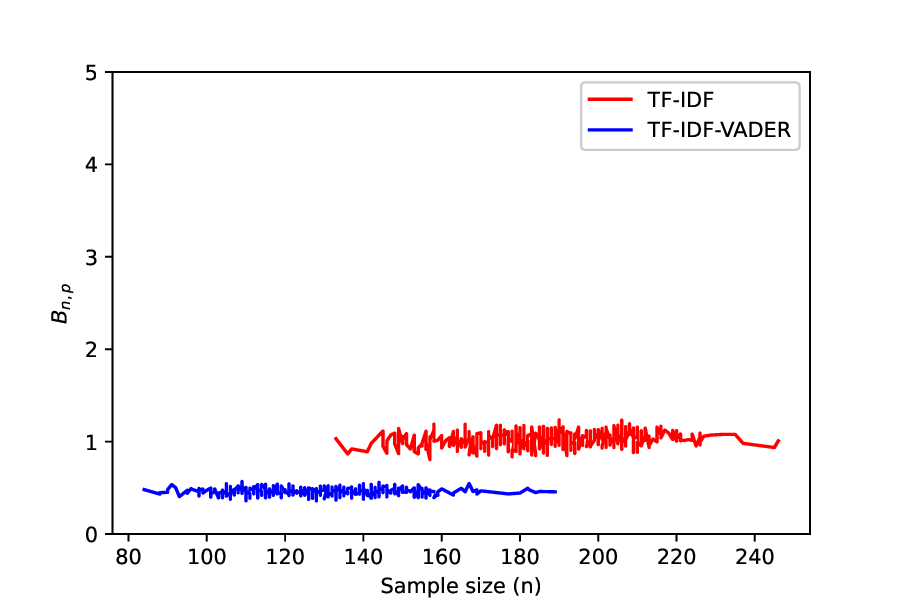}}  
\subfigure[$p=0.9^{-1}$] {\includegraphics[width=0.49\textwidth]{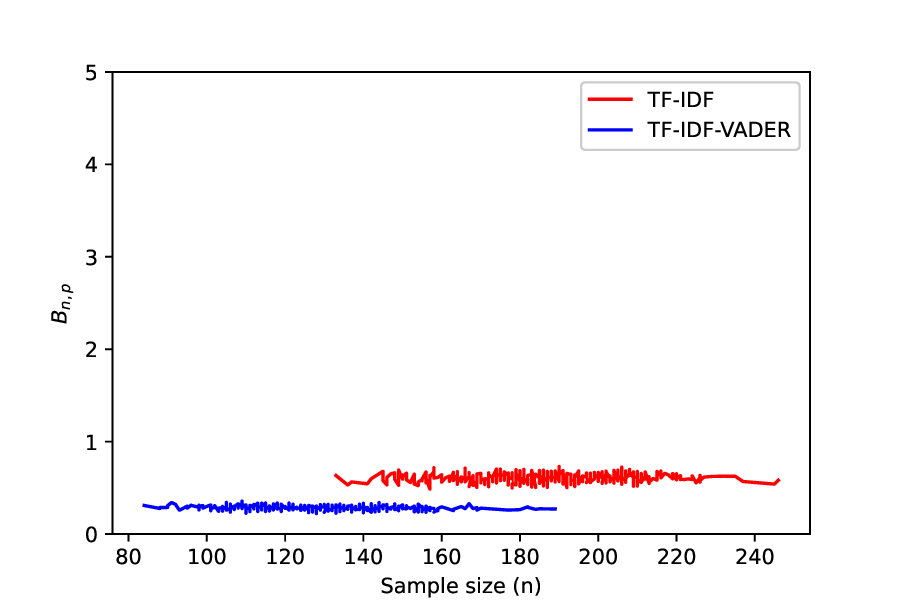}}
\caption{The plots of $B_{n,p}$ (vertical axis) for various sample sizes $n$ (horizontal axis) and parameter $p$ values  when using the GB classification.}
\label{fig:Bnp GB}
\end{figure}

\begin{figure}[h!]
\centering
\subfigure[$p=0.6^{-1}$] {\includegraphics[width=0.49\textwidth]{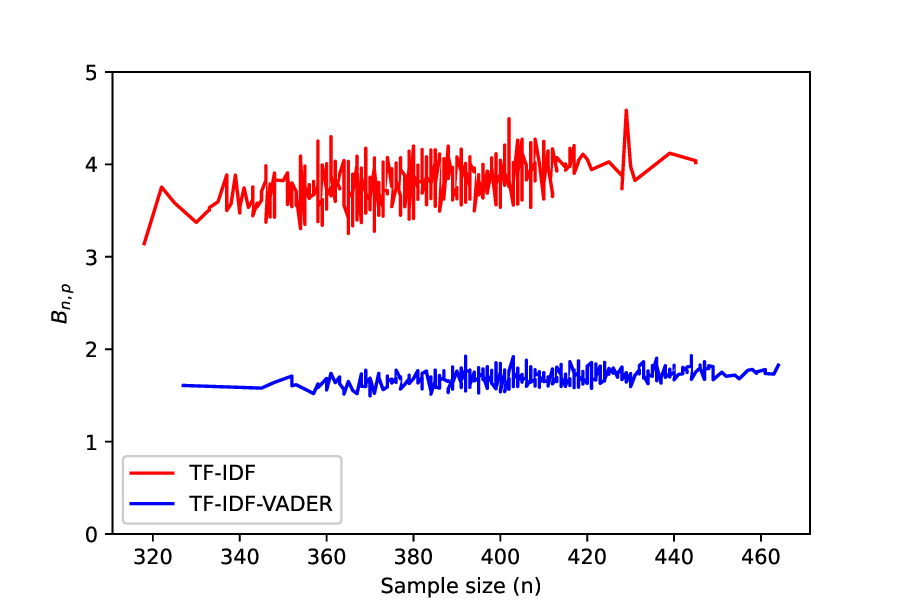}}  
\subfigure[$p=0.7^{-1}$] {\includegraphics[width=0.49\textwidth]{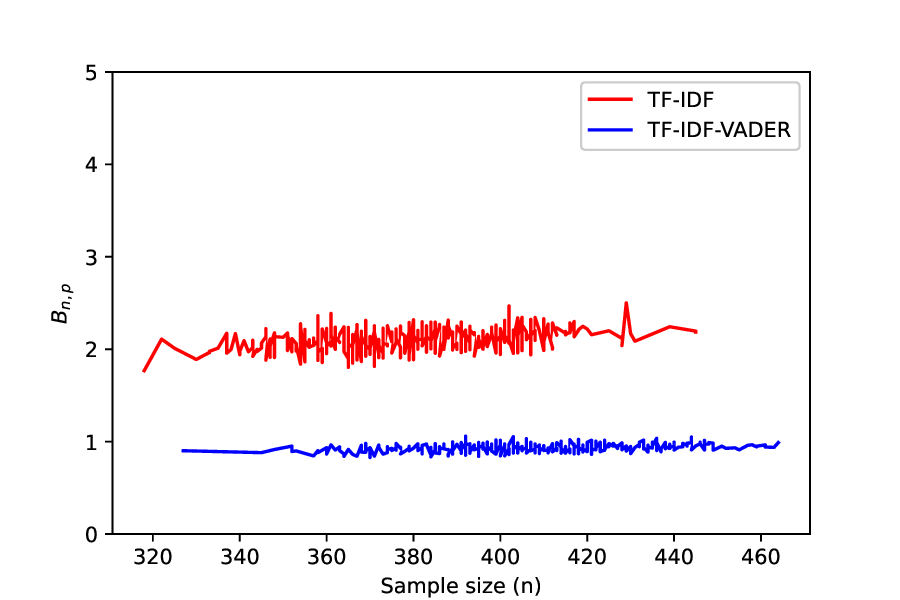}}
\\
\subfigure[$p=0.8^{-1}$] {\includegraphics[width=0.49\textwidth]{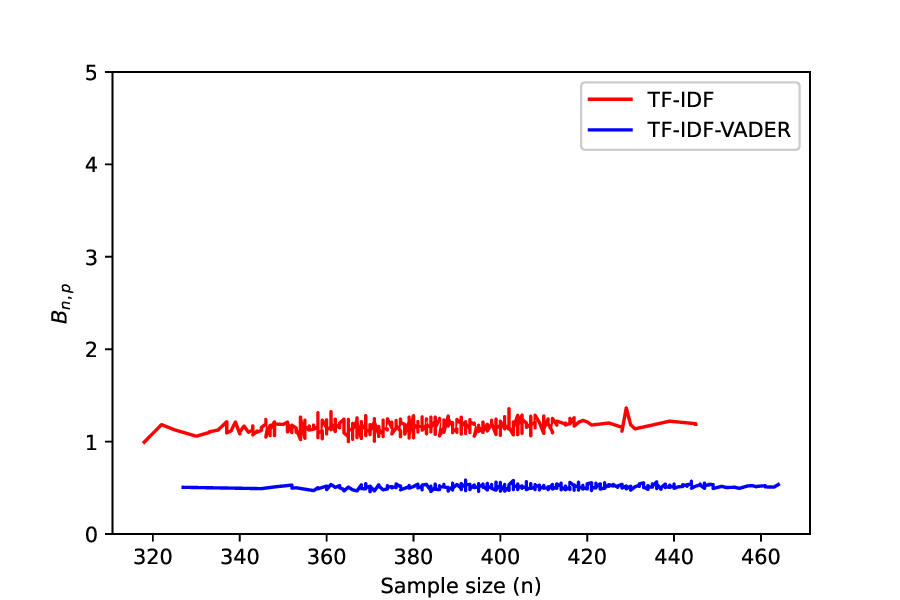}}  
\subfigure[$p=0.9^{-1}$] {\includegraphics[width=0.49\textwidth]{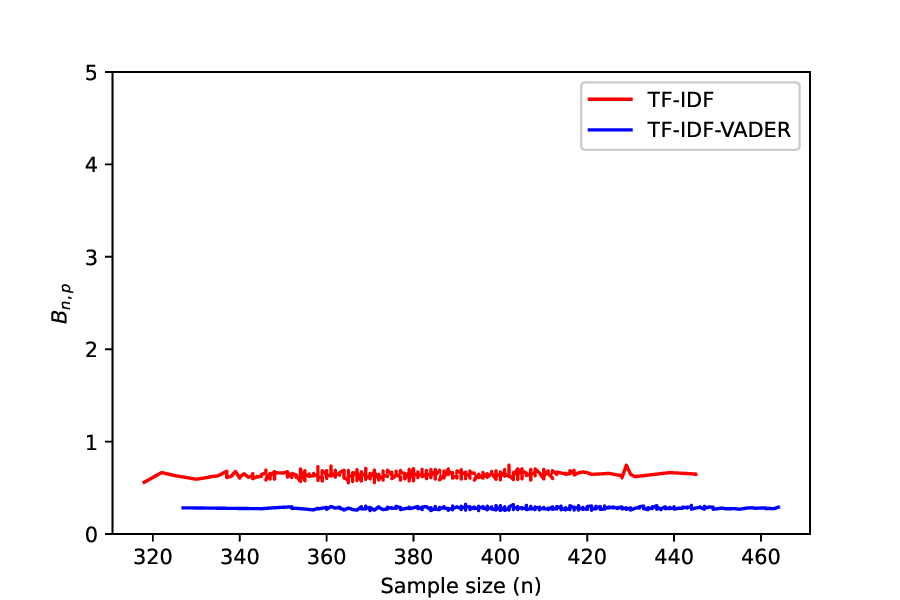}}
\caption{The plots of $B_{n,p}$ (vertical axis) for various sample sizes $n$ (horizontal axis) and parameter $p$ values when using the MLP classification.}
\label{fig:Bnp MLP}
\end{figure}

\clearpage 
\newpage 

\begin{figure}[h!]
\centering
\subfigure[$p=0.6^{-1}$] {\includegraphics[width=0.49\textwidth]{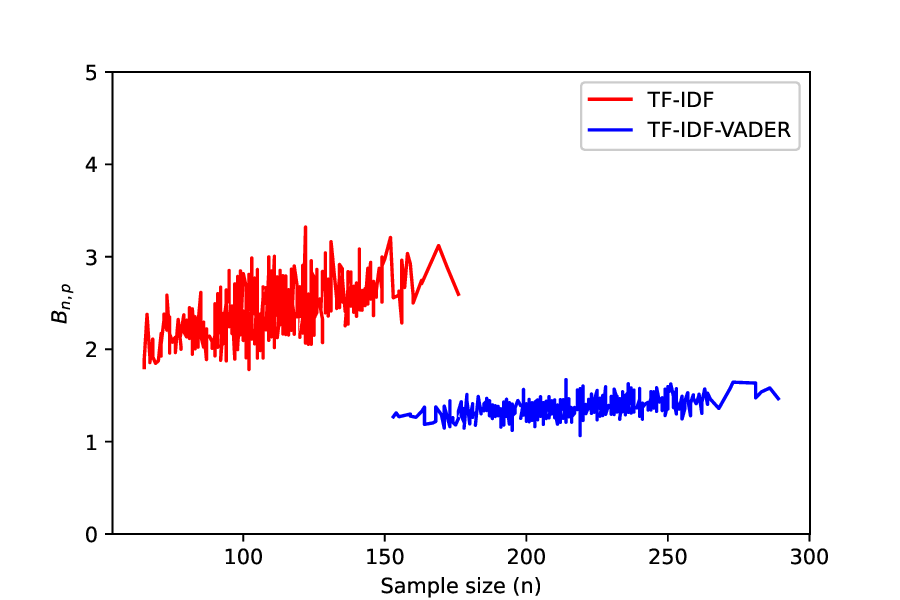}}  
\subfigure[$p=0.7^{-1}$] {\includegraphics[width=0.49\textwidth]{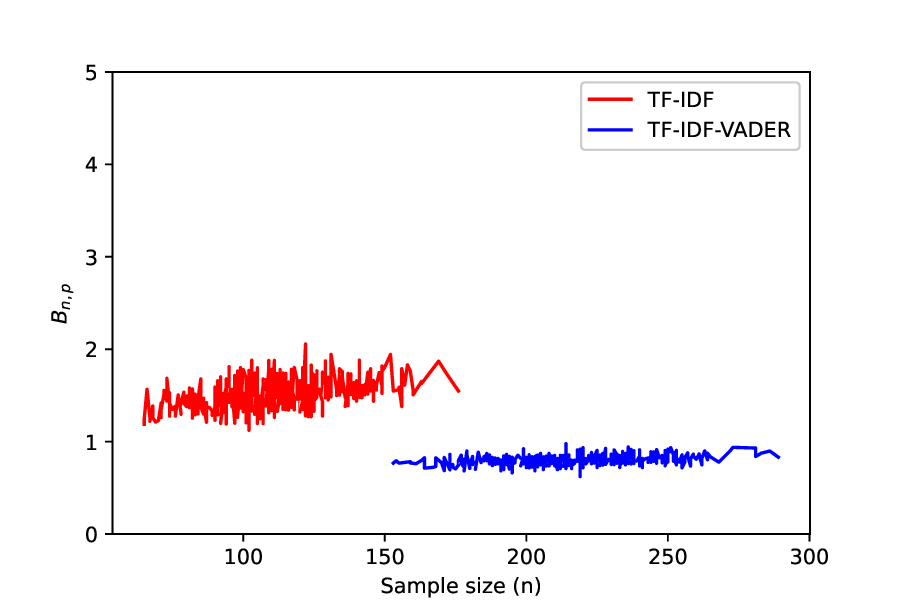}}
\\
\subfigure[$p=0.8^{-1}$] {\includegraphics[width=0.49\textwidth]{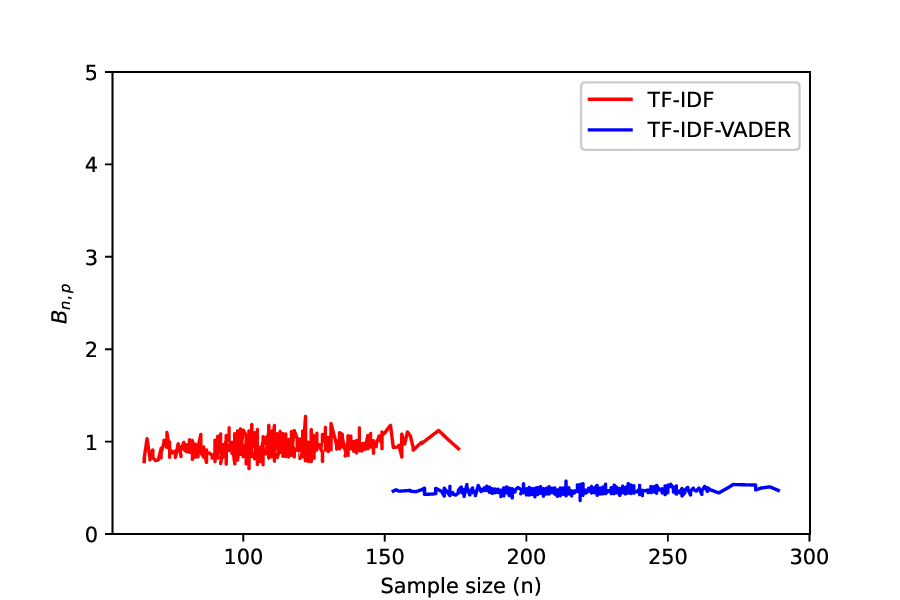}}  
\subfigure[$p=0.9^{-1}$] {\includegraphics[width=0.49\textwidth]{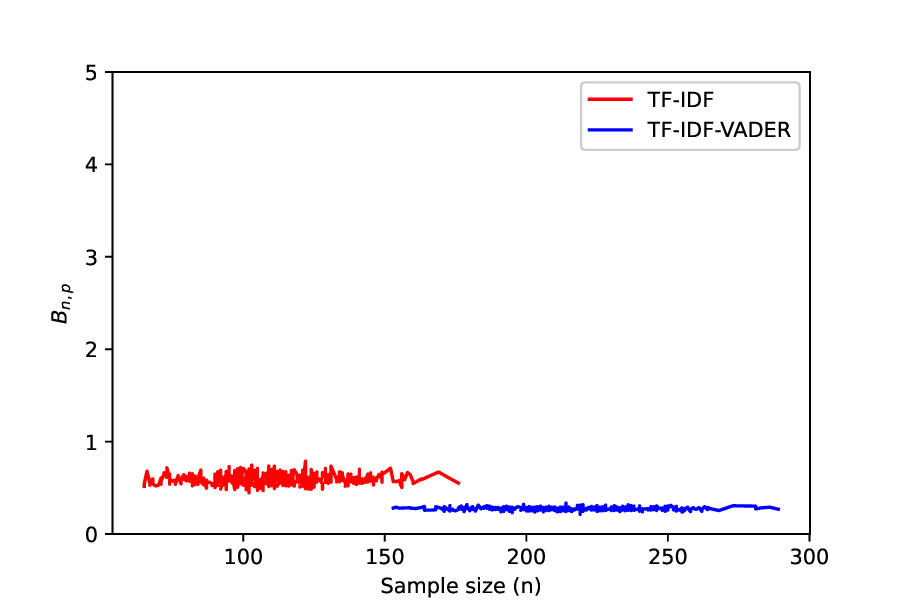}}
\caption{The plots of $B_{n,p}$ (vertical axis) for various sample sizes $n$ (horizontal axis) and parameter $p$ values when using the RF classification.}
\label{fig:Bnp RF}
\end{figure}

\end{document}